\documentclass[journal]{IEEEtran}
\makeatother
\usepackage{fonttable}
\usepackage{amsmath}
\usepackage{amsfonts}
\usepackage{graphicx} 
\usepackage{url}
\usepackage{subfigure}
\usepackage{amsmath}
\usepackage{amsbsy}
\usepackage{algorithm}
\usepackage{algpseudocode}
\usepackage{amsmath, amssymb, bm}
\usepackage{amssymb}
\usepackage{graphicx,xparse}
\usepackage{mathtools}
\usepackage{placeins}
\usepackage{pdfpages}
\usepackage{mathptmx}
\usepackage{graphicx}
\usepackage{subfigure}
\usepackage{amsmath}
\usepackage{amsbsy}
\usepackage{amssymb}
\usepackage{graphicx,xparse}
\usepackage{mathtools}
\usepackage{subcaption}
\usepackage{mathrsfs} 
\usepackage{algpseudocode}
\usepackage{multirow}
\usepackage{array}
\usepackage{caption}
\usepackage[utf8]{inputenc}
\usepackage{acronym}
\usepackage{mathptmx}
\usepackage{enumitem}
\setlist[itemize]{noitemsep, topsep=0pt}
\setlist[enumerate]{noitemsep, topsep=0pt}
\usepackage{longtable}
\usepackage[hidelinks]{hyperref}
\usepackage{xcolor}
\hypersetup{
    colorlinks,
    linkcolor={black!50!black},
    citecolor={black!90!black},
    urlcolor={black!80!black}
}
\usepackage{xcolor}
\usepackage{cite}
\usepackage{comment}
\usepackage[acronym]{glossaries}

\newacronym{6g}{6G}{sixth generation}
\newacronym{ris}{RIS}{reconfigurable intelligent surfaces}
\newacronym{dnn}{DNN}{Deep Neural Network}
\newacronym{ann}{ANN}{Artificial Neural Network}
\newacronym{los}{LoS}{line-of-sight}
\newacronym{so}{SOs}{scattering objects}
\newacronym{bilstm}{biLSTM}{bidirectional long-short term memory}
\newacronym{lstm}{LSTM}{long-short term memory}
\newacronym{ue}{UE}{user equipment}
\newacronym{rmse}{RMSE}{root mean squared error}
\newacronym{bs}{BS}{base station}
\newacronym{snr}{SNR}{signal-to-noise ratio}
\newacronym{rse}{RSE}{rich-scattering environment}
\newacronym{rses}{RSEs}{rich-scattering environments}
\newacronym{dl}{DL}{deep learning}
\newacronym{ml}{ML}{machine learning}
\newacronym{rnn}{RNN}{recurrent neural network}
\newacronym{mimo}{MIMO}{multiple-input-multiple-output}
\newacronym{siso}{SISO}{single-input-single-output}
\newacronym{crlb}{CRLB}{Cramér-Rao lower bound}
\newacronym{bcrlb}{BCRLB}{Bayesian Cramér-Rao lower bound}
\newacronym{ofdm}{OFDM}{frequency division
multiplexing}
\newacronym{rss}{RSS}{received signal strength}
\newacronym{miso}{MISO}{multiple-input-single-output}
\newacronym{bo}{BO}{Bayesian Optimization}
\newacronym{bisen}{Bi-SEN}{Scattering Estimation Network}
\newacronym{biaruln}{Bi-ARULN}{Adaptive RIS-Assisted User Localization Network}
\newacronym{cdf}{CDF}{cumulative distribution function}
\newacronym{aoa}{AOA}{angle-of-arrival}
\newacronym{fim}{FIM}{Fisher Information Matrix}

\title{Adaptive RIS Configuration Design with Environmental Sensing for User Localization in Dynamic Rich Scattering Environment}
\author{Anum~Umer, Ivo~M\"{u}\"{u}rsepp, Muhammad~Mahtab~Alam,~\IEEEmembership{Senior~Member,~IEEE}
\thanks{A. Umer, I. M\"{u}\"{u}rsepp, and M. M. Alam are with Thomas Johann Seebeck Department of Electronics, Tallinn University of Technology, 12616 Tallinn, Estonia. E-mail: \{anum.umer, ivo.muursepp, muhammad.alam\}@taltech.ee.} 
\thanks{This work was supported in part by the European Union’s Horizon 2020 Research and Innovation Program under Grant 101058505 ‘5G-TIMBER’; in part by the Project ``Increasing the Knowledge Intensity of Ida-Viru Entrepreneurship'' co-funded by the European Union; and in part by the NATO-SPS G7699 PROTECT Project. This work was also supported by Telia Eesti AS through TalTech Development Fund.}}

\begin{document}
\maketitle

%

\begin{abstract} 
 This paper addresses the problem of adaptive \acrfull{ris} configuration design for user localization in \acrfull{rse}, where electromagnetic waves undergo multiple interactions with dynamic scatterers and \acrshort{ris} elements. We propose an adaptive learning-based localization approach for a distributed \acrshort{ris}-assisted network in a \acrshort{rse} using a \acrfull{bilstm} model that captures temporal correlations between observations. The proposed approach actively senses the environment using sequential pilot transmissions from the \acrfull{bs}, accounting for scattering effects, and adaptively updates the \acrshort{ris} configuration based on prior measurements to eventually accurately estimate and minimize the user localization error. The proposed model comprises two neural sub-networks: \acrfull{bisen}, for estimation of scattering in the environment, and \acrfull{biaruln}, for RIS configuration and localization. Bayesian optimization is used for hyperparameter tuning of the model. The simulation results demonstrate the effectiveness of the proposed approach, achieving significantly lower localization \acrfull{rmse} compared to random configuration, prestored codebook look-ups, and adaptive baselines in both \acrfull{siso} and \acrfull{mimo} \acrshort{ris}-assisted networks in \acrshort{rse}. The design is generalized across configurations and scales with \acrshort{ris} size and network dimensions. The results highlight the strong potential of \acrshort{ris} deployment and of the proposed approach to enable reliable location services in \acrshort{rse}.
\end{abstract}


\section{Introduction}

Traditional localization techniques often fail in rich multipath or rapidly changing environments-herein referred to as \acrfull{rses}-and face infrastructure limitations, particularly indoors \cite{9847080,bourdoux20206g}. The programmable nature of \acrfull{ris} offers a transformative solution by enabling fine-grained control over the wireless propagation environment with minimal power consumption \cite{9140329, PhysRevApplied.11.044024,li2017electromagnetic}. Composed of numerous passive elements that manipulate the phase of reflected electromagnetic waves, RIS can act as intelligent anchors or improve existing infrastructure to overcome challenges such as synchronization, interference, and directional ambiguity \cite{9847080,9548046,bourdoux20206g}. These capabilities make RIS especially valuable for precise, sustainable, and adaptive localization—critical for applications such as autonomous vehicles, robotics, virtual reality, smart factories, and traffic systems \cite{10044963, bourdoux20206g, 10858311}. With 6G’s all-spectrum integration-from microwave to terahertz and optical bands-indoor environments like offices and industrial settings are particularly prone to such rich scattering effects \cite{you2021towards}. 

Dynamic \acrshort{rse} introduces complex propagation conditions, where electromagnetic waves experience multiple interactions with surrounding objects and RIS elements, resulting in a superposition of reflected signals with varying \acrfull{aoa} and polarizations \cite{9856592,10453467,10133065}. Unlike free space, where wave propagation is predictable and RIS parameterization is typically linear (via \acrfull{bs}-\acrshort{ris}-receiver cascades), or quasi-free space, where RIS serves as an alternative relay due to blocked \acrfull{los} paths, RSE conditions lead to highly nonlinear and dynamic RIS behavior \cite{10858311}. Here, RIS act as controlled perturbations, while the environment contributes uncontrolled, time-varying disturbances such as moving objects or humans-herein referred to as \acrfull{so}, resulting in a nonlinear double-parameterized system influenced by both RIS configurations and environmental factors \cite{10077120}. Addressing these challenges requires advanced RIS configuration algorithms that transcend free-space assumptions and leverage environmental context to minimize localization errors and meet the precision and robustness demands of next-generation localization, sensing, and communication systems \cite{10352433,10453467,10858311}. This requires context-aware adaptation of RIS, achieved through sensing and estimating the evolving state of \acrshort{so}. Although incorporating sensing capabilities into \acrshort{ris} offers a promising direction \cite{10352433}, the optimal design of RIS configurations guided by such environmental awareness in dynamic \acrshort{rse} remains an open research challenge.

This work considers the problem of accurate user localization in a dynamic \acrshort{rse} using a distributed \acrshort{ris}-assisted network deployed in a reverberant enclosure. In the considered scenario, simulated with a physics-based end-to-end channel \cite{9856592}, a \acrshort{bs}, a user, and multiple non-cooperative \acrshort{so}, each following their own fixed trajectory, are present. The BS transmits pilot signals, while the RIS assists in redirecting them toward the user. A subset of RIS elements is dedicated to actively sensing the environment to estimate the scattering contributed by \acrshort{so} and quantized by their location estimate. Estimation of \acrshort{so} location at any instant can leverage RIS-assisted sensing by utilizing a predefined sequence of randomized RIS configurations. This is achieved using hybrid meta-atoms \cite{10352433}, which simultaneously capture and reflect electromagnetic waves, providing sufficient configurational diversity to enable the extraction of noncooperative \acrshort{so} information from transmission measurements between a single node pair \cite{9860667}. Learned sensing further enhances this capability by optimizing RIS configuration sequences to improve the accuracy of user localization \cite{9593167}. However, the underlying relationships between RIS configurations, user measurements, and scattering are highly nonlinear and sequential in nature. As measurements accumulate, the need for scalable, context-aware solutions becomes essential. We address this by sequentially configuring the RIS using an adaptive, data-driven model based on prior and current measurements. This dynamic sensing and configuration mechanism enables the system to track evolving multipath conditions and minimize localization error without relying on static configurations or pre-stored mappings.

To address the challenges of user localization in dynamic \acrshort{rse}, we propose a deep learning (\acrshort{dl})-based approach that jointly performs  \acrshort{ris} configuration design and user position estimation. Leveraging prior measurements and real-time sensing of \acrshort{so}, the RIS is adaptively configured to improve localization accuracy. At the core of our approach is a recurrent neural network with \acrfull{bilstm} units, which capture long-term dependencies and temporal patterns in sequential observations by processing data in both forward and backward directions \cite{10.1162/neco.1997.9.8.1735}. This enables the dynamic construction of compact information vectors for guiding RIS updates and refining user location estimates without relying on complex real-time computations or static pre-stored mappings. The \acrshort{bilstm} architecture is particularly suited for \acrshort{rse} environments due to its adaptability to diverse data types, efficient training through parallelization, and superior performance over traditional \acrfull{ml} models that rely on manual feature extraction \cite{10225284,9255727}. By integrating \acrshort{dl} with RIS-assisted \acrshort{rse}, the system continuously learns to extract key signal characteristics and environmental features, yielding accurate and robust localization in complex propagation scenarios \cite{9898733}. The numerical results demonstrate superior localization performance in \acrfull{siso} and \acrfull{mimo} \acrshort{rse} scenarios, highlighting the generalizability and effectiveness of the proposed approach.

\subsection{Related Work}
Numerous recent studies have explored RIS-assisted localization, focusing on fingerprinting-based methods, optimization and error-bound driven approaches, and more recent learning-based solutions, all aiming to enhance localization accuracy by exploiting RIS reconfigurability
\cite{9548046,9124848,9500437,9500281,9500663,9540372,9193909,9456027,10054103,10373816,10549836,9508872,10740619}. The authors in \cite{9548046} enhance diversity in fingerprint-based localization by leveraging the \acrshort{ris} ability to reconfigure its electromagnetic response by selecting optimal configurations from a large codebook to generate distinct radio maps. This variation in configurations improves the discriminability of \acrfull{rss} fingerprints, enriching the feature space and improving localization accuracy without requiring multiple access points. In \cite{9124848} a hierarchical RIS codebook is proposed to facilitate adaptive bisection search across the angular space for enhanced 2D localization, although such codebooks may be suboptimal in low \acrfull{snr} conditions and training overhead can be substantial \cite{9414523}. In \cite{9500663}, the authors categorize RIS profiles into three distinct classes and analyze their effects on the 3D positioning error bound of the user. Two classes are designed heuristically to generate narrow or broad beams, while the third class comprises randomly generated RIS profiles. The authors in \cite{9193909} and \cite{9456027} improve indoor localization by optimizing RIS profiles to maximize \acrfull{rss} variations between nearby locations, using local and global search techniques. Their proposed scheme achieves a threefold reduction in localization error compared to traditional \acrshort{rss} based methods. Analytically derived profiles, as in \cite{10054103}, jointly optimize communication throughput and localization performance based on geometric models, but often rely on idealized channel assumptions that may not hold in practice for \acrshort{rses}.

Other studies focus on channel-aware optimization and theoretical error bounds. A codebook-based \acrshort{ris} configuration method for user localization in \acrshort{ris}-assisted \acrshort{siso} networks under \acrshort{rse} is proposed in \cite{10765779}, leveraging historical and real-time channel data to jointly predict optimal \acrshort{ris} settings and user position. The approach achieves up to 97.16\% reduction in localization error compared to random \acrshort{ris} configurations. In \cite{9500437} a gradient descent method is used to iteratively adjust \acrshort{ris} reflection coefficients to minimize the \acrfull{crlb} of the user location. In \cite{9508872}, the CRLB is derived for near-field localization and orientation in \acrshort{ris}-assisted systems using synchronous and asynchronous signaling, and a closed-form \acrshort{ris} phase profile is proposed to maximize \acrshort{snr} at the BS while accounting for spherical wavefronts. Similarly, \cite{9148744} investigates \acrshort{ris}-aided downlink positioning by optimizing \acrshort{ris} elements to improve the \acrfull{fim} rank, emphasizing the importance of directing the reflected energy toward the user. In \cite{9977919}, a multi-\acrshort{ris}-assisted millimeter positioning system is examined, where \acrshort{ris} phase shifts are optimized using particle swarm optimization to minimize the position and rotation error bounds. Channel estimation and user positioning in \acrshort{ris}-assisted \acrshort{mimo}-\acrfull{ofdm} systems are studied in \cite{9540372}, which proposes four \acrshort{ris} training coefficient designs—random, structured, grouping, and sparse—for improved estimation followed by \acrshort{aoa}-based positioning. These studies demonstrate that \acrshort{crlb}-based formulations can provide strong theoretical guidance. Extensions such as Bayesian \acrshort{fim} and robust \acrfull{bcrlb} allow uncertainty in prior knowledge to be formally integrated, for example, through prior distributions on channel or geometry parameters \cite{6541985}. Adaptive \acrshort{ris} beamforming strategies that exploit the estimated location of the user and associated uncertainty have been proposed in \cite{9772371}, although these designs neglect scattering effects, which may limit their applicability in dynamic \acrshort{rses}.

Learning-based approaches for sequential beamforming and \acrshort{ris} optimization have also been investigated. In \cite{9414523} and \cite{8792366}, sequential beamforming for initial access in single-path millimeter channels is addressed using a \acrfull{dnn} that maps the estimated \acrshort{aoa} posterior distribution to the next beamforming vector. This is extended in \cite{9724252} to multipath channel models settings using an \acrshort{lstm} model, while \cite{9746028} employs Gated Recurrent Units for the same task. The \acrshort{lstm} in \cite{9724252} is also applied to beam tracking, where \acrshort{ris} is dynamically adjusted to follow user movement over time \cite{10097134}. A sequential \acrshort{ris} and BS beamforming design is proposed in \cite{10279094} for \acrshort{ris}-assisted localization in a \acrshort{siso} network using \acrshort{lstm}, and extended to a multi-\acrshort{ris} \acrfull{miso} network in \cite{10373816}. An \acrshort{lstm}-based approach is also proposed in \cite{10549836} to optimize user positioning through adaptive \acrshort{ris} control and power allocation, enhancing channel capacity. These methods highlight the potential of data-driven solutions to exploit temporal correlation in user mobility. However, these works do not address adaptive \acrshort{ris} configuration design for user localization as a function of previous measurements while accounting for the impact of scattering in \acrshort{rses}.


\subsection{Main Contributions}
In this study, we introduce a learning-based user localization approach for a \acrshort{ris}-assisted \acrshort{rse} that dynamically adjusts the \acrshort{ris} configuration based on sensing of scattering in the environment, contributed by dynamic \acrshort{so}, as well as historical environmental measurements to improve accuracy. Unlike free-space localization \cite{9215972}, which relies on ray-tracing techniques and \acrshort{aoa} estimation, rich-scattering conditions involve multiple reflections from various angles, rendering conventional methods ineffective \cite{10077120}. Instead, we extract information about object location and movement from scattered wave patterns by learning the correlation between \acrshort{so} positions and received field measurements. Our approach leverages a data-driven optimization strategy to refine the mapping from prior observations to future \acrshort{ris} configurations, enabling accurate localization. In such environments, signal propagation exhibits long-range dependencies due to scattering, which \acrshort{bilstm} effectively models through sequential data processing. This enables the model to adapt to dynamic channel and \acrshort{ris} conditions while mitigating gradient vanishing or explosion, ensuring robust performance \cite{10.1162/neco.1997.9.8.1735}.  At each time step, a \acrshort{bilstm} cell processes current and past measurements to update a hidden state that represents the environment. This hidden state undergoes successive updates through multiple layers until it effectively encodes the necessary information for designing the next \acrshort{ris} configuration. After accumulating sufficient measurements, the final cell state is passed through a fully connected neural network, which generates the estimated position of the user. The approach supports both moving \acrshort{so} and user scenarios, with \acrshort{ris} configurational diversity enabling efficient sensing at a single frequency with minimal sensing nodes. The main contributions of this work are as follows.  
\begin{enumerate}
    \item We propose an adaptive sensing-based localization approach for a distributed \acrshort{ris}-assisted \acrshort{siso} network in \acrshort{rse}, where \acrshort{ris} configurations are dynamically designed using prior measurements and sensed scattering in the environment. The designed \acrshort{ris} configuration is used to enable better user location estimates in terms of accuracy.

    \item We employ a \acrshort{bilstm} model to map pilot signal responses, from user and \acrshort{ris} elements dedicated to environment sensing, to \acrshort{ris} configurations. The model consists of two sequential networks: \acrshort{bisen} (\textit{\acrlong{bisen}}) for scattering estimation, feeding context into \acrshort{biaruln} (\textit{\acrlong{biaruln}}) for \acrshort{ris} configuration design and user localization. Additionally, \acrfull{bo} is integrated for hyperparameter tuning \cite{hutter2019automated, NIPS2012_05311655, Feurer2019} to improve prediction accuracy by efficiently identifying optimal model settings.


    \item We extend the proposed approach to a \acrshort{mimo} network to evaluate the impact of antenna array size at transceiver nodes in dynamic \acrshort{rse} scenarios.
\end{enumerate}
 The proposed approach produces interpretable results, demonstrating that the sequential design of the \acrshort{ris} configuration progressively enhances the \acrshort{snr} of received pilots and improves localization accuracy. Simulation results confirm that the adaptive \acrshort{ris} configuration design in \acrshort{rse} via the devised approach achieves lower localization errors than conventional random configuration, non-adaptive pre-stored configuration lookup based \acrshort{ris} designs, or adaptive designs not taking scattering into account. When extended to a \acrshort{mimo} network, the proposed learning-based adaptive approach delivers a higher localization accuracy compared to the \acrshort{siso} network, demonstrating the advantages of having multiple antenna nodes in addition to \acrshort{ris} in \acrshort{rses}.  

\subsection{Paper Organization and Notations}
The remainder of the paper is organized as follows. The system model and the formulation of the problem are presented in Section \ref{systemmodel}. Section \ref{method} presents the architecture of the devised approach. The results are presented in Section \ref{results}. Conclusions are derived in Section \ref{conclusion}.

\textit{Notations}: Scalars, vectors, and matrices are represented in italic (e.g. \( a \)), lowercase bold letters (e.g. \( \mathbf{a} \)), and capital boldface letters (e.g. \( \mathbf{A} \)), respectively. Furthermore, \ \( |\cdot| \) denotes the modulus,  \( \hat{a} \) denotes an estimate of \( a \), $\text{diag}(\mathbf{x})$ presents a diagonal matrix with entries of $\mathbf{x}$ on diagonal.  The real and imaginary components of a complex number are represented by \( \Re(\cdot) \) and \( \Im(\cdot) \), respectively. The expectation of a random variable is denoted by \( \mathbb{E}(\cdot) \). The complex Gaussian distribution with mean $m$ and variance $v$ is presented by $\mathcal{CN}(m,v)$.

\section{System Model and Problem Formulation}
\label{systemmodel}
\subsection{System Model}
\begin{figure}
    \centering
    \includegraphics[width=1\linewidth]{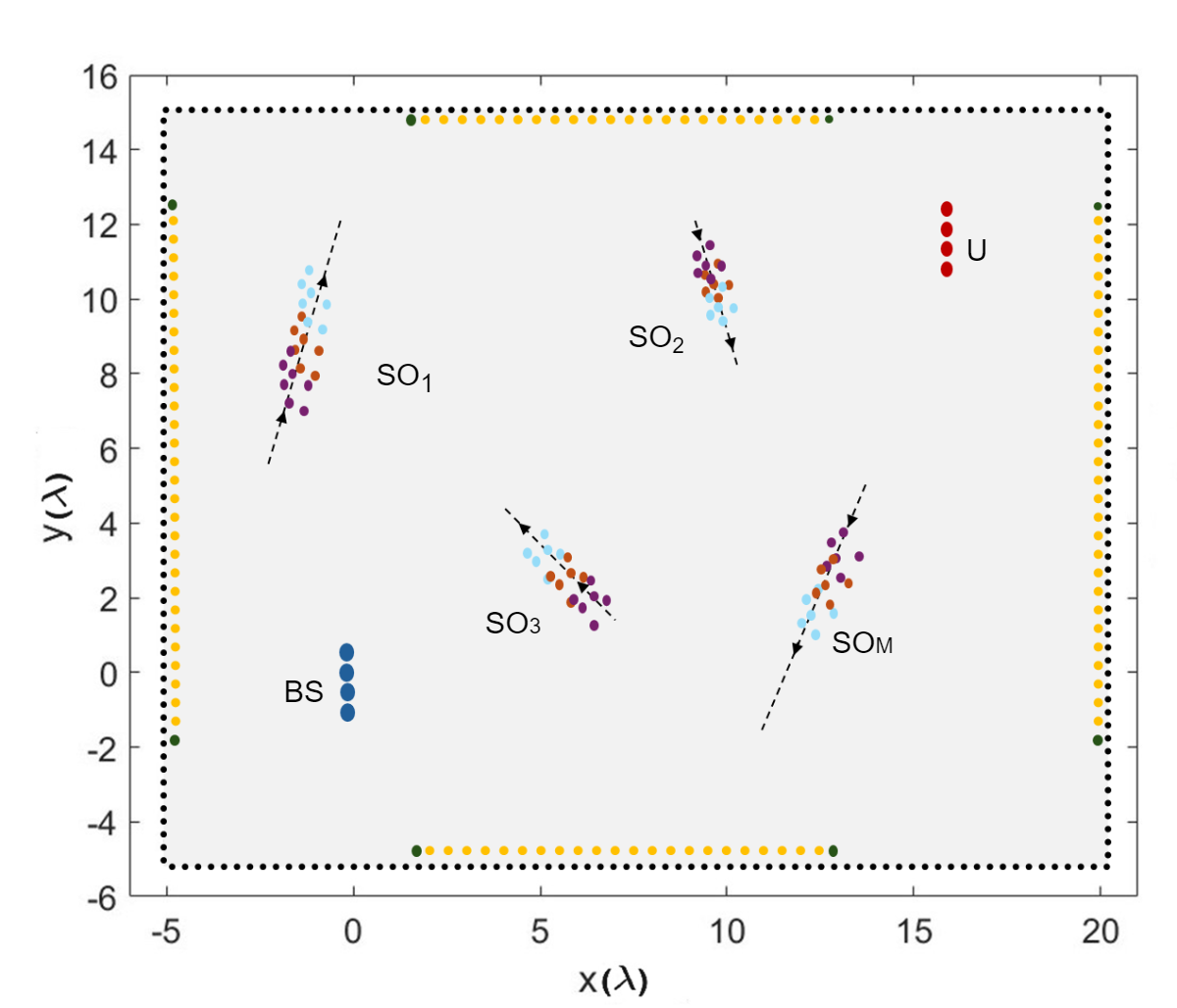}
    \caption{Illustration of considered \acrshort{rse}, modeled in physics-based end-to-end channel \cite{9856592} as a many-wavelengths large regularly shaped room (described by wavelength $\lambda$ scale). The \acrshort{rse} consists of $N_E$ dipoles whose locations are indicated by black dipole fence and collections of clusters of colored dipoles at various locations, inside the enclosure, presenting \acrshort{so}. The location of dipoles presented with black color always stays the same, while the location of colored dipoles is varied randomly, in each simulation realization, along their fixed trajectories indicated by the dotted line. The locations of the dynamic dipoles presenting \acrshort{so} is shown in three different colors for first three realizations (plum, orange, cyan corresponding to each realization). The \acrshort{rse} is equipped with $N_{\text{BS}}$ antenna BS (blue dipoles), a $N_{\text{U}}$ antenna user equipment $U$ (red dipoles), and a uniform linear array of distributed \acrshort{ris} installation with $N_{\text{RIS}}$ elements (yellow dipoles) that partially covers the four walls (black dipole fence) of the enclosure. $S_{\text{RIS}}$ denotes the subset of \acrshort{ris} elements (green dipoles) used by the \acrshort{bs} to estimate the \acrshort{so} location status based on the measurement of the wireless channel from $S_{\text{RIS}}$.} 
    \label{fig1}
\end{figure}

This paper considers the problem of localization in a distributed \acrshort{ris}-assisted enclosed indoor scenario, presenting a \acrshort{rse}, where a user equipment \(U\) with $N_{\text{U}}$ antennas aims to estimate its location $\mathbf{u}$ with the help of a \acrshort{bs} equipped with $N_{\text{BS}}$ antennas and uniform linear array of \acrshort{ris} strategically placed on the periphery of the communication environment as in Figure \ref{fig1} (top view). To maintain tractability and relevance to structured \acrshort{rses} such as industrial or controlled indoor settings, we assume that there are $M$ \acrshort{so} present in the environment and each of them moves along its fixed trajectory. For simplicity, the \acrshort{so} positions are updated simultaneously, while their movements remain independent along their respective paths. 


The distributed \acrshort{ris} installation configuration design is managed by a \acrshort{ris} controller, which operates based on control signals received from \acrshort{bs}. It comprises of \(N_{\text{RIS}}\) reflective elements, each applying a tunable phase shift to the incident signal. The configuration of these elements is represented by the vector \(\mathbf{k} \in \mathbb{C}^{1 \times N_{\text{RIS}}}\), where each entry satisfies the unit-modulus constraint, i.e., \(|[\mathbf{k}]_n| = 1, \forall n \in \{1, \dots, N_{\text{RIS}}\}\). This ensures that each element reflects the incoming wave with a controllable phase shift \(\delta_n \in [0, 2\pi)\) while preserving its amplitude. In addition, \(S_{\text{RIS}}\), a subset of RIS elements placed at fixed locations, is specifically allocated for channel sensing where the power splitting ratio is set to zero, resulting in no signal reflection \cite{10352433, 10077120}.


Electromagnetic waves undergo multiple reflections and scatter off from walls. \acrshort{ris} and \acrshort{so}, causing significant signal scattering, while the movement of \acrshort{so} further alters propagation paths, creating a dynamic and complex fading wireless communication environment\cite{9475155}. The estimated location status of $M$ \acrshort{so}, via \(S_{\text{RIS}}\), is represented by the vector $\mathbf{p} \in \mathbb{R}^{2M}$.

\begin{figure}
\centering
\includegraphics[width=1\linewidth]{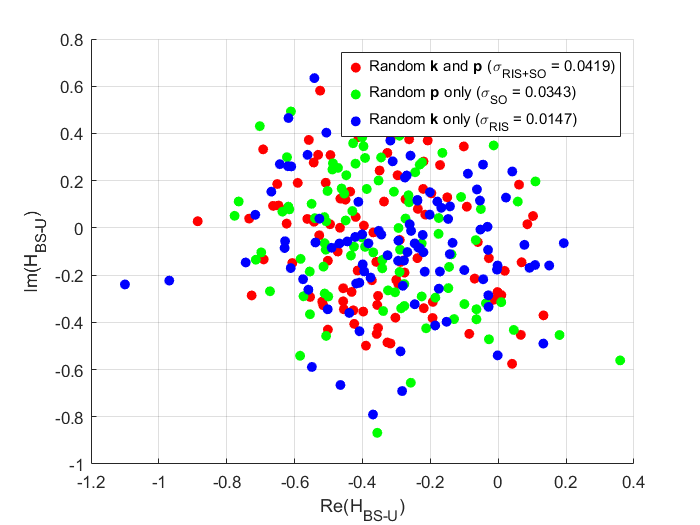}
    \caption{
    Distribution of $\textbf{H}_{\text{BS}-\text{U}}$ in the complex plane for a \acrshort{siso} setup, shown over 100 random realizations of \acrshort{ris} configurations $\mathbf{k}$ and \acrshort{so} locations $\mathbf{p}$ (red), of $\mathbf{p}$ with fixed $\mathbf{k}$ (green), and of $\mathbf{k}$ with fixed $\mathbf{p}$ (blue), simulated with \cite{9856592}. The SO positions are sampled uniformly along their predefined trajectories (as shown in Figure \ref{fig1}), and \acrshort{ris} phase configurations are sampled independently and uniformly from the range $[0, 2\pi)$. Here, $\sigma$ is the standard deviation of the complex-valued $\textbf{H}_{\text{BS}-\text{U}}$.
    }
    \label{fig2}
\end{figure}

\subsection{End-to-End Channel Modeling}
We assume that the channel is constant within a coherence period. In \acrshort{rse}, subtle, configuration-dependent fluctuations in observed signal responses may occur due to fine-grained environmental perturbations, e.g., slight \acrshort{so} motion or multipath dynamics. These fluctuations are modeled as structured fast-fading behavior in discrete time. This allows our model to capture temporal dependencies across RIS configurations for improved scattering and user location estimation.
Let \( \textbf{H}_{\text{BS}-\text{U}} \in \mathbb{C}^{N_{\text{U}} \times N_{\text{BS}}} \) represent the stochastic channel frequency response at the user, encompassing both the \acrshort{los} signals and multipath components that interact with the RIS elements, \acrshort{so}, and the walls. 

The influence of \acrshort{ris} configuration $\mathbf{k}$ and moving \acrshort{so} location $\mathbf{p}$ on statistical properties of the channel response at the user, $\textbf{H}_{\text{BS-U}}$, simulated with the physics-based end-to-end channel model \cite{9856592}, is illustrated in Figure~\ref{fig2}. It shows the standard deviation $\sigma$ of the complex-valued $\textbf{H}_{\text{BS-U}}$ distribution when $\mathbf{k}$, $\mathbf{p}$, or both are varied independently. As expected, when $\mathbf{k}$ and $\mathbf{p}$ are changed simultaneously, the standard deviation increases. These observations underscore the significant parametric dependence of $\textbf{H}_{\text{BS-U}}$ on both $\mathbf{k}$ and $\mathbf{p}$. This renders simplified analytical models, typically used in free-space scenarios, inadequate for \acrshort{rse}. In free-space conditions, the channel \( \textbf{H}_{\text{BS}-\text{U}} \) is typically modeled as a linear cascade from the \acrshort{bs} to the \acrshort{ris} and then to the user. In contrast, \acrshort{rse} introduces complex multipath interactions involving multiple \acrshort{ris} elements and dynamic \acrshort{so}, making the channel highly nonlinear and analytically intractable \cite{9475155}.


To address this, we simulate the \acrshort{rse} using a physics-based end-to-end channel simulator specifically designed for \acrshort{ris}-parameterized environments\cite{9856592}. Based on the coupled-dipole formalism, end-to-end channel models wireless entities as dipole assemblies with configurable properties such as resonance and absorption, enabling realistic simulation of scattering effects and ensuring compliance with fundamental electromagnetic principles.
\\
As shown in Figure \ref{fig1}, end-to-end channel models  \acrshort{so}, walls, \acrshort{ris} and transceiver surfaces as dipole arrays. Note that $N\triangleq N_\text{BS}+ N_\text{U} + N_\text{E} + N_{\text{RIS}}$ gives the total count of dipoles within the system where $N_\text{E}$ dipoles constitute the \acrshort{rse} and $N_\text{RIS}$ dipoles constitute the \acrshort{ris}.  To simplify the analysis, the model operates in a two-dimensional space, with dipoles positioned in the \( x \)-\( y \) plane with their moments aligned along the \( z \)-axis.

The channel frequency response at the user is given by
\begin{equation}
\mathbf{H}_{\text{BS}-\text{U}} = \mathbf{V}\left[(N_{\text{BS}} + 1) : (N_{\text{BS}} + N_{\text{U}}), 1 : N_{\text{BS}} \right],   
\end{equation}
where
\begin{equation}
\mathbf{V} \triangleq \text{diag} \left( \left[ \alpha^{-1}_1, \alpha^{-1}_2, \dots, \alpha^{-1}_N \right] \right) \mathbf{W}^{-1},
\end{equation}
and $\alpha^{-1}_i$ is the inverse frequency-dependent polarizability of the $i^{th}$ dipole. 
The \( i^{th} \) diagonal element of \( N \times N \) complex-valued matrix \(\mathbf W \) represents the inverse polarizability \( \alpha^{-1}_i (f) \) of the \( i^{th} \) dipole.  The polarizability \( \alpha_i (f) \) of the \( i \)th dipole is given by
\begin{equation}
\alpha_i(f) = \frac{\chi^2_i}{4\pi^2 f^2_{\text{res},i} - 4\pi^2 f^2 + j(\gamma^R_i + 2\pi f \Gamma^L_i)},
\end{equation}
where \( \chi^2_i \) represents a charge-related term functioning as an amplitude factor for \( \alpha_i(f) \). The parameter \( f_{\text{res},i} \) denotes the resonance frequency, while \( \gamma^R_i \) accounts for intrinsic radiation damping. Additionally, \( \Gamma^L_i \geq 0 \) corresponds to the absorptive damping term.

The off-diagonal entry in \(\mathbf W \) at position \( (i, j) \) is given by \( -G_{ij} (f) \), which corresponds to the negative of the two-dimensional free-space Green’s function describing the interaction between the \( i \)th and \( j \)th dipoles at
the positions $\mathbf{r}_i$ and $\mathbf{r}_j$, given by
\begin{equation}
G_{ij} (\mathbf{r}_i, \mathbf{r}_j, f) \triangleq -j \frac{k^2}{4\epsilon\delta} \text{H}^{(2)}_0 \left( k |r_i - r_j| \right).
\end{equation}
where \( \text{H}^{(2)}_0 (\cdot) \) denotes a Hankel function of the second kind.

The channel response from the end-to-end channel modeling accurately incorporates aspects such as space-time causality, frequency-selective dispersion, phase-amplitude coupling of \acrshort{ris} elements, mutual coupling effects, and long-range mesoscopic correlations, providing a realistic depiction of wave propagation in complex scattering environments.  

\subsection{Signal Transmission and Reception}
Assuming a narrowband system model, the BS transmits a sequence of \( T \) known pilot symbols \( \textbf{x}^{(t)} \in \mathbb{C}^{N_{BS} \times 1} \) to the user over \( T \) time frames when a localization request is initiated. The received signal  \( \textbf{y}^{(t)} \in \mathbb{C}^{N_{U} \times 1}\) at the user includes contributions from the direct path, if available, and multipath components reflected by the \acrshort{ris} and scattering environment. We assume that the BS receives the user's responses through a low-rate feedback link, enabling centralized inference for \acrshort{ris} configuration and localization. The input-output relationship is given by:  
\begin{equation}
\textbf{y}^{(t)} = \textbf{H}_{\text{BS}-\text{U}}^{(t)}  \textbf{x}^{(t)} + \textbf{w}^{(t)}, \quad t \in \mathbb{Z}
\end{equation}
where \( \textbf{w}^{(t)} \sim  \mathcal{CN}(0,\sigma^2 )\) is additive Gaussian noise.   

In parallel, the BS uses \( S_{\text{RIS}} \) dedicated \acrshort{ris} elements set to sense the environment by observing the channel \( \mathbf{H}_{\text{BS}-\text{S}}  \in \mathbb{C}^{{\text{S}_{\text{RIS}}} \times N_{\text{BS}}}\), between \acrshort{bs} and \( S_{\text{RIS}} \) with observations $\textbf{y}_{s}^{(t)} = \textbf{H}_{\text{BS}-\text{S}}^{(t)} \mathbf{x}^{(t)} + \textbf{w}^{(t)}, \quad t \in \mathbb{Z}$. This sensing enables inference of the scattering object (SO) location status \( \mathbf{p} \) at an instant $t$, providing an estimate of the scattering in \acrshort{rse} without requiring cooperation from the user.    

Unlike conventional methods that rely on random \acrshort{ris} configurations, this paper aims to adaptively optimize the \acrshort{ris} configuration $\mathbf{k}$ to improve user localization accuracy in \acrshort{rse}. The proposed approach assumes a centralized architecture in which the BS coordinates the scattering estimation, \acrshort{ris} configuration updates and user localization.

\subsection{Localization in RSE}
The goal of the localization task is to estimate the unknown position of the user \( \mathbf{u} \) from \( T \) received observations \( [\mathbf{y}^{(t)}]_{t=0}^{T-1} \), given the known positions of the \acrshort{bs} and \acrshort{ris} and predicted \acrshort{so} location. The \acrshort{so} location status \( \mathbf{p} \) is inferred from \acrshort{ris} sensing signals \( [\mathbf{y}_s^{(\tau)}]_{\tau=0}^{t} \) collected by \( S_{\text{RIS}} \) elements, to account for environmental scattering in the \acrshort{rse}. In this setting, \acrshort{ris} serves as a cost-effective source of directional information, acting as a set of virtual anchors through its dense and distributed reflective elements. Although \acrshort{ris} cannot transmit or receive directly, it aids in localization by dynamically adjusting its configuration to induce observable variations in the received signal. 

Previous studies suggest that strategically selecting \acrshort{ris} configurations in \acrshort{rse} can improve the localization accuracy by shaping the received signal distribution to make nearby spatial samples yield more distinguishable observations \cite{10765779,9860667,ref9}. However, most studies rely on fixed configurations that remain unchanged during operation, using predetermined beams to scan the search area at set angles. In contrast, this paper introduces an adaptive sensing-based approach that sequentially refines the search space by dynamically updating \acrshort{ris} configurations based on received pilot signals and sensed environmental changes, thus improving the accuracy of localization.

\acrshort{ris} operation in \acrshort{rse} involves two key stages: sensing the environment to gain context awareness and identifying the optimal configurations to support user localization. This process requires offline training tailored to the specific \acrshort{rse} and online inference during deployment. Sensing focuses on estimating the \acrshort{so} state from field measurements, enabling informed configuration design for the subsequent localization process.

In a \acrshort{rse}, the observed field at any point results from the superposition of multipath signals, with variations introduced by dynamic \acrshort{so} depending on their location \( \mathbf{p} \). The channel response \( \textbf{H}_{\text{BS}-\text{U}} \) reflects the combined effects of the \acrshort{ris} configurations and \acrshort{so} interactions on the transmitted pilot signal, while \( \textbf{H}_{\text{BS}-\text{S}} \) captures information used to infer \( \mathbf{p} \). A sequence of such measurements forms a unique wave fingerprint that enables user localization.

To adapt to temporal variations in scattering, the \acrshort{so} location \( \hat{\mathbf{p}}^{(t)} \) at time \( t \) is estimated as a function of the pilot signals sensed at $S_{\text{RIS}}$, 
\begin{equation}
\label{pt}
    \hat{\mathbf{p}}^{(t)} = G^{(t)}([\textbf{y}_s^{(\tau)}]_{\tau=0}^{t}]),
\end{equation}
where \( G^{(t)}(\cdot):\mathbb{C}^{t+1} \to \mathbb{R}^{2M}\) represents mapping from previously sensed pilots to the estimated  \acrshort{so} location. Here, $\tau$ accounts for previous time steps up to the current time $t$.

Using the estimate in \eqref{pt} and the user's prior received observations, i.e., $[\textbf{y}^{(\tau)}]_{\tau=0}^{t}$, the BS determines the next set of $\mathbf{k}$ \acrshort{ris} configurations, denoted as $\textbf{k}^{(t+1)}$, at time $t$. The \acrshort{ris} configuration vector at the $(t+1)$-th stage is expressed as 
\begin{equation}
\label{F}
\textbf{k}^{(t+1)} = F^{(t)}\left( [\textbf{y}^{(\tau)}]_{\tau=0}^{t}, \hat{\mathbf{p}}^{(t)} \right),
\end{equation}
where $F^{(t)}: \mathbb{C}^{t+1} \times \mathbb{R}^{2M}\to  \mathbb{C}^{1 \times N_{\text{RIS}}}$ defines the mapping from the previously received pilot signals and estimated \acrshort{so} location to the next \acrshort{ris} configuration. The updated \acrshort{ris} configuration is then used to generate the next measurement $\textbf{y}^{(t+1)}$ in time frame $(t+1)$.  For initialization at $t < 0$, there are no prior observations. Here, the function $F^{(-1)}(\emptyset)$ takes an empty input set and provides a fixed default initial \acrshort{ris} configurations $\mathbf{k}^{(0)}$.

After collecting all $T$ pilot signals, the estimated user location $\hat{\mathbf{u}}^{(T)}$ is obtained as a function of all past observations at the user and environment sensing:
\begin{equation}
\label{Q}
\hat{\mathbf{u}}^{(T)} = Q\left( [\textbf{y}^{(t)}]_{t=0}^{T-1},  \hat{\mathbf{p}}^{(t)}\right),
\end{equation}
where $Q: \mathbb{C}^T \times \mathbb{R}^{2M} \to \mathbb{R}^2$ maps the collected pilot signals and the estimated scattering to the estimated two-dimensional user location.

The general objective is to minimize localization error by jointly optimizing \( \{G^{(t)}\}_{t=0}^{T-1} \) in \eqref{pt}, \( \{F^{(t)}\}_{t=0}^{T-1} \) in \eqref{F} and \( Q \) in \eqref{Q}, under the following constraints:
\begin{subequations} \label{eq:optimization_problem}
\begin{align}
\min_{\{G^{(t)}(\cdot), F^{(t)}(\cdot)\}_{t=0}^{T-1}, Q(\cdot)} & \sqrt{\mathbb{E} \left[ \|\hat{\mathbf{u}}^{(T)} - \mathbf{u}\|^2 \right]} \label{eq:optimization_objective} \\
\text{subject to}
& \label{9b}\quad \hat{\mathbf{p}}^{(t)} = G^{(t)}([\textbf{y}_s^{(\tau)}]_{\tau=0}^{t})\\
&\quad  \mathbf{k}^{(t+1)}  = F^{(t)}\left( [\textbf{y}^{(\tau)}]_{\tau=0}^{t}, \hat{\mathbf{p}}^{(t)} \right), \label{9c} \\
& \quad \hat{\mathbf{u}}^{(T)} = Q\left( [\textbf{y}(t)]_{t=0}^{T-1}, \hat{\mathbf{p}}^{(t)} \right). \label{9d}\\
& \quad \left| \left[ \mathbf{k}^{(t+1)} \right]_n \right| = 1, \quad \forall n \in N_{RIS}, t. \label{9e}
\end{align}
\end{subequations}
The solution of the optimization problem in \eqref{eq:optimization_problem} is challenging due to the joint estimation of scattering \( \{G^{(t)}\}_{t=0}^{T-1} \), the \acrshort{ris} configuration design $\{F^{(t)}(\cdot)\}_{t=0}^{T-1}$ and user localization $Q(\cdot)$. To simplify the problem, a common approach is to use codebooks to select \acrshort{ris} configurations heuristically. For example,  \cite{9124848} addresses \acrshort{ris}-aided localization in free space by employing a hierarchical codebook to adaptively scan the space. While codebooks are appealing due to their low control overhead and hardware feasibility, such strategies inherently limit design flexibility by restricting \acrshort{ris} configurations to a discrete predefined set, which cannot fully adapt to the dynamic nature of \acrshort{rse}, thereby constraining localization accuracy.

In this paper, we directly optimize \eqref{eq:optimization_problem} using a neural network to parameterize the functions \( G^{(t)}(\cdot) \), \( F^{(t)}(\cdot) \) and \( Q(\cdot) \). This enables scalable adaptation to rich scattering conditions while maintaining localization accuracy and computational efficiency. As demonstrated in Section~\ref{results}, our learning-based strategy significantly outperforms codebook-based designs in dynamic RSE environments.

\section{Proposed Adaptive biLSTM based Localization Approach}
\label{method}
We present a deep learning-based adaptive \acrshort{ris} configuration approach to address the localization problem in \eqref{eq:optimization_problem}, without relying on a predefined geometric channel model or restricting the design space to a fixed codebook. At each time step \( t \), the RIS configuration \( \textbf{k}^{(t+1)} \) is determined based on prior user observations \( [\textbf{y}^{(\tau)}]_{\tau=0}^{t} \) and the estimated SO status \( \mathbf{p}^{(t)} \).  A core challenge is the linear accumulation of measurements over time, making it computationally expensive to process them when optimizing \acrshort{ris} configurations for large \( t \). To maintain scalability, the neural network must condense past observations into a compact fixed-dimensional information vector.


To this end, we employ a \acrshort{bilstm} model that captures temporal dependencies and efficiently encodes sequential measurements. Unlike conventional approaches that adapt \acrshort{ris} configurations solely based on user pilot signals, the proposed method introduces a two-stage deep learning model:
\begin{enumerate}
    \item  \textit{\acrfull{bisen}}: Estimates dynamic \acrshort{so} positions at time \( t \) using sensing observations at $S_{\text{RIS}}$, characterizing the current scattering in \acrshort{rse}.
    \item \textit{\acrfull{biaruln}}: Designs the RIS configurations for the time slot $(t+1)$ based on the estimated scattering via \acrshort{bisen} and pilot received at the user at $t$, with the objective of localizing the user at final time $T$.
\end{enumerate}
By leveraging a bidirectional recurrent neural network, the proposed approach captures dependencies from both past and future contexts, improving scattering estimation and \acrshort{ris} adaptation for user localization.

\subsection{Bayesian Optimization Based Hyperparameter Tuning}
We employ \acrshort{bo} to fine-tune the hyperparameters of the \acrshort{bisen} and \acrshort{biaruln} models prior to training, efficiently identifying optimal settings for the number of BiLSTM layers $N_L$, hidden state dimension $d_s$, learning rate $\eta$ and batch size $B$. This optimally balances computational efficiency and localization accuracy. \acrshort{bo} models the relationship between hyperparameters and validation loss using a Gaussian Process (GP) regression model,
\begin{equation}
L_{\text{val}}(N_L, d_s, \eta, B) \sim GP(\mu(\Theta), k(\Theta, \Theta'))
\end{equation}
where $L_{\text{val}}$ is the validation loss, $\mu(\Theta)$ is the mean function and $k(\Theta, \Theta')$ is the covariance kernel function measuring the similarity between hyperparameter sets \( \Theta = (N_L, d_s, \eta, B) \).

To explore promising hyperparameter configurations, \acrshort{bo} uses the Expected Improvement (EI) acquisition function,
\begin{equation}
\alpha_{\text{EI}}(\Theta) = \mathbb{E}[\max(0, L_{\min} - L_{\text{val}}(\Theta))].
\end{equation}
where, \( L_{\min} \) denotes the best validation loss observed among all previously evaluated hyperparameter configurations. The next set of hyperparameters is selected as
\begin{equation}
\Theta^* = \arg\max_{\Theta} \alpha_{\text{EI}}(\Theta).
\end{equation}

This process enables effective hyperparameter tuning of both models while minimizing overall computational overhead.

\begin{figure*}
    \centering   \includegraphics[width=1\linewidth]{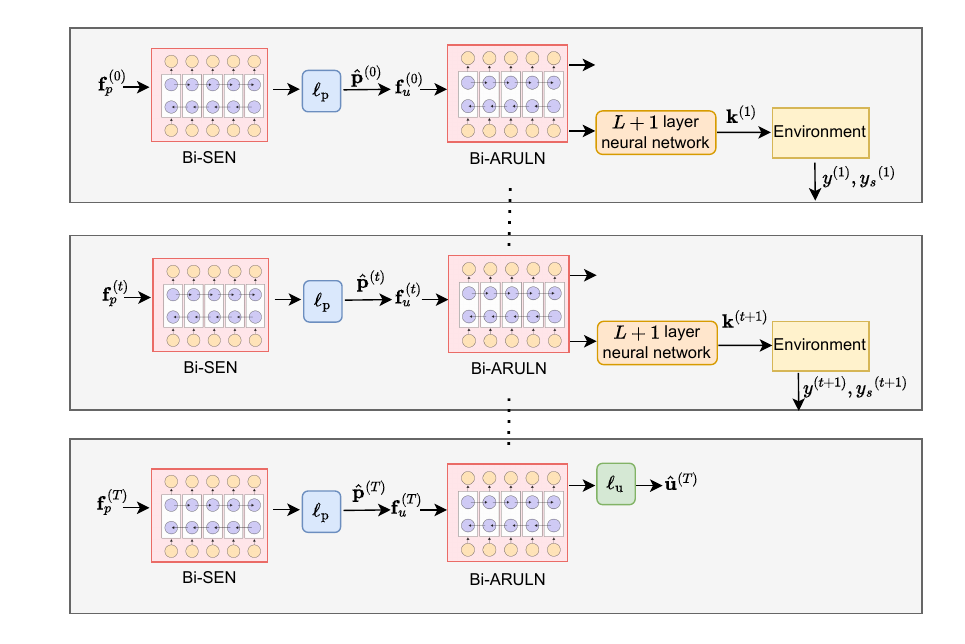}
    \caption{Proposed adaptive localization approach for \acrshort{ris}-assisted \acrshort{rse}. At each time step \( t \in [0, T-1] \), sensed signals from $S_{\text{RIS}}$ is fed as input feature set $\mathbf{f}_p^{(t)} = [ |\textbf{y}_s^{(t)}|, \angle{\textbf{y}_s^{(t)}}]$ to \acrshort{bisen} to estimate \acrshort{so} locations \( \hat{\mathbf{p}}^{(t)} \), corresponding to (\ref{9b}). The feature set $\mathbf{f}_u^{(t)} = [ |\textbf{y}^{(t)}|, \angle {\textbf{y}^{(t)}}, \hat{\mathbf{p}}^{(t)}]$ is input to \acrshort{biaruln}, which predicts the next \acrshort{ris} configuration \( \mathbf{k}^{(t+1)} \), corresponding to (\ref{9c}) and (\ref{9e}). The designed configuration \( \mathbf{k}^{(t+1)} \) is used to make next received pilot measurements $\textbf{y}_s^{(t+1)}$ and $\textbf{y}^{(t+1)}$  in $(t+1)$-th time frame. After all the pilot measurements have been received, user location \( \hat{\mathbf{u}}^{(T)} \) is estimated at the $T$-th time frame via a feedforward prediction layer, as a function of all $T$ historical measurements, corresponding to (\ref{9d}). }
    \label{nnarch}
\end{figure*}

\subsection{Neural Network Architecture}
\label{arch}
The network architecture is illustrated in Figure \ref{nnarch}. The proposed BiLSTM-based localization model comprises two sequential networks-a \acrshort{bisen} and \acrshort{biaruln}. Both networks incorporate fully connected layers for feature extraction and BiLSTM layers for processing sequential data across multiple sensing frames.
\subsubsection{Bi-SEN Network}
At each time step $t$,  a \acrshort{bilstm} cell updates its forward- and backward hidden states, $\bm{s}_{p,f}^{(t)}$ and $\bm{s}_{p,b}^{(t)}$, and  forward- and backward cell states $\bm{c}_{p,f}^{(t)}$ and $\bm{c}_{p,b}^{(t)}$, based on the newly received input \( \mathbf{f}_p^{(t)} \). The input vector is formed by concatenating the magnitude and phase of the sensed signal, i.e., $\mathbf{f}_p^{(t)} = [ |\textbf{y}_s^{(t)}|, \angle{\textbf{y}_s^{(t)}} ]$. 

The updates for the forward and backward cell states and hidden states are given by
\begin{subequations}
\label{eq:bilstm}
\begin{align}
    \bm{c}_{p,f}^{(t)} &= \bm{f}_{p,f}^{(t)} \circ \bm{c}_{p,f}^{(t-1)} + \bm{i}_{p,f}^{(t)} \circ \tanh ( \bm{u}_{cp} (\mathbf{f}_p^{(t)}) + \bm{r}_{cp} (\bm{s}_{p,f}^{(t-1)},\bm{s}_{p,b}^{(t+1)}) ),\label{cfp}\\
    \bm{s}_{p,f}^{(t)} &= \bm{o}_{p,f}^{(t)} \circ \tanh(\bm{c}_{p,f}^{(t)}),\label{sfp}\\
    \bm{c}_{p,b}^{(t)} &= \bm{f}_{p,b}^{(t)} \circ \bm{c}_{p,b}^{(t+1)} + \bm{i}_{p,b}^{(t)} \circ \tanh(\bm{u}_{cp}(\mathbf{f}_p^{(t)}) + \bm{r}_{cp}(\bm{s}_{p,f}^{(t-1)}, \bm{s}_{p,b}^{(t+1)})),\label{cbp}\\
    \bm{s}_{p,b}^{(t)} &= \bm{o}_{p,b}^{(t)} \circ \tanh(\bm{c}_{p,b}^{(t)}),\label{sbp}
\end{align}    
\end{subequations}
where $\bm{u}_{cp}(\cdot)$ and $\bm{r}_{cp}(\cdot)$ are fully connected linear layers, and $\bm{f}_{p,.}^{(t)}$, $\bm{i}_{p,.}^{(t)}$, and $\bm{o}_{p,.}^{(t)}$ represent the forget, input, and output gate activation vectors, respectively. 
The element-wise updates for these gates are defined as
\begin{subequations}
\label{gates}
\begin{align}
\bm{f}_{p,f}^{(t)} = \sigma_s(\bm{u}_{p,f}(\mathbf{f}_p^{(t)}) + \bm{r}_{p,f}(\bm{s}_{p,f}^{(t-1)})),\label{forget}\\
\bm{i}_{p,f}^{(t)} = \sigma_s(\bm{u}_{p,i}(\mathbf{f}_p^{(t)}) + \bm{r}_{p,i}(\bm{s}_{p,f}^{(t-1)})),\label{input}\\
\bm{o}_{p,f}^{(t)} = \sigma_s(\bm{u}_{p,o}(\mathbf{f}_p^{(t)}) + \bm{r}_{p,o}(\bm{s}_{p,f}^{(t-1)})),\label{output}
\end{align}
\end{subequations}
where the element-wise sigmoid function is $\sigma(x) = \frac{1}{1 + e^{-x}}$. The parameters $\bm{u}_{p,f}, \bm{u}_{p,i}, \bm{u}_{p,o}, \bm{r}_{p,f}, \bm{r}_{p,i}, \bm{r}_{p,o}$ are trainable linear layers in the network. The backward states update follows the similar structure to ensure bidirectional processing.

The final estimated \acrshort{so} positions are computed via a fully connected network output layer
\begin{equation}
\label{scatestim}
\hat{\mathbf{p}}^{(t)} = \ell_{\text{p}}([c_{p,f}^{(t)};c_{p,b}^{(t)}]),
\end{equation}
where \( \ell_{\text{p}}(\cdot) \) is a trainable fully connected network. The estimate in (\ref{scatestim}) corresponds to the criterion in (\ref{9b}).

\subsubsection{Bi-ARULN Network}
 A \acrshort{bilstm} cell processes newly received input to update its forward and backward hidden states, denoted as $\bm{s}_{u,f}^{(t)}$ and $\bm{s}_{u,b}^{(t)}$, along with the corresponding cell states $\bm{c}_{u,f}^{(t)}$ and $\bm{c}_{u,b}^{(t)}$. The input vector at time $t$, $\mathbf{f}_u^{(t)}$, consists of the magnitude and phase of the pilots received at the user, and the estimated $\hat{\mathbf{p}}^{(t)}$, given by $\mathbf{f}_u^{(t)} = [ |\textbf{y}^{(t)}|, \angle {\textbf{y}^{(t)}}, \hat{\mathbf{p}}^{(t)} ]$. The update equations for the forward and backward cells, hidden states, and their respective gates are as in \eqref{eq:bilstm} and \eqref{gates}.

The final hidden state at time $t$ is obtained by concatenating the forward and backward hidden states,
\begin{equation}
\label{st}
s_u(t) = [s_{u,f}(t); s_{u,b}(t)].
\end{equation} 

The hidden state vector $s_u(t)$ is used as input to a fully connected neural network with $L_n$ layers to derive the next \acrshort{ris} configurations at an instant $t+1$. The transformation is performed as
\begin{equation}
\label{gamma}
\gamma(t) = \beta_{L_{n}} \left( \bm{W}_{L_{n}} \beta_{L_{n}-1} \left( \dots \beta_1 \left( \bm{W}_1 s_u(t) + \bm{b}_1 \right) \dots \right) + \bm{b}_{L_{n}} \right),
\end{equation}
where $\beta_l$ is the activation function for the $l$-th layer, set as $\text{ReLU}(x) = \max(0, x)$ where $l \in \{1, \dots, L_n\}$. The parameters $\bm{W}_l$ and $\bm{b}_l$ are trainable weights and biases.

The $(L_n+1)$-th layer outputs the real-valued vector
\begin{equation}
\bar{\mathbf{k}}^{(t+1)} = \dot{\bm{W}}_{L_{n}+1} \gamma(t) + \dot{\bm{b}}_{L_{n}+1}.
\end{equation}

Here, $\bar{\mathbf{k}}^{(t+1)}$ represents the sequence of real and imaginary components of the \acrshort{ris} configuration vector. The trainable weight and bias matrices, \( \dot{\bm{W}}_{L_{n}+1} \), \( \dot{\bm{b}}_{L_{n}+1} \), are structured to maintain the appropriate output dimensions such that \( \bar{\mathbf{k}}^{(t+1)} \in \mathbb{R}^{2N_{\text{RIS}}} \), producing \( N_{\text{RIS}} \) complex-valued entries. To ensure that each element of the \acrshort{ris} configuration vector satisfies the unit-modulus constraint, it undergoes element-wise normalization:
\begin{equation}
\label{risconfig}
\begin{split}
[\mathbf{k}^{(t+1)}]_n =
\frac{[\Re(\bar{\mathbf{k}}^{(t+1)})]_n}{\sqrt{[\Re(\bar{\mathbf{k}}^{(t+1)})]_n^2 + [\Im(\bar{\mathbf{k}}^{(t+1)})]_n^2}}\\
+ j \frac{[\Im(\bar{\mathbf{k}}^{(t+1)})]_n}{\sqrt{[\Re(\bar{\mathbf{k}}^{(t+1)})]_n^2 + \Im[(\bar{\mathbf{k}}^{(t+1)})]_n^2}},
\end{split}
\end{equation}
Design in (\ref{risconfig}) corresponds to the criterion in (\ref{9c}) and (\ref{9e}), respectively. 

The final estimated \acrshort{ue} position is obtained in the time frame \( T \) as 
\begin{equation}
\label{estimuser}
    \hat{\mathbf{u}}^{(T)} = \ell_u([c_{u,f}^{(T)}; c_{u,b}^{(T)}]),
\end{equation}
where \( \ell_u(\cdot) \) is a fully connected network that processes the concatenated forward and backward cell states to obtain the final position estimate as in (\ref{9d}).

\subsection{Loss Function}
The overall BiLSTM model is trained using the Adam optimizer \cite{Kingma2014AdamAM} to minimize the \acrfull{rmse} loss between the true and estimated user location. The loss for user localization in \acrshort{biaruln}, corresponding to (\ref{eq:optimization_objective}), is defined as
\begin{equation}
\label{lossu}
    \sqrt{\mathbb{E} \left[ \|\hat{\mathbf{u}}^{(T)} - \mathbf{u}\|^2 \right]}.
\end{equation}

To account for scattering estimation loss, the cumulative \acrshort{rmse} loss in localizing $M$ \acrshort{so} with \acrshort{bisen} at each stage is given by
\begin{equation}
\label{lossp}
    \sqrt{ \mathbb{E} \left[ \|\hat{\mathbf{p}}^{(t)} - \mathbf{p}\|^2 \right]}.
\end{equation}
These losses guide the model to accurately estimate location of the user and adapt to environmental changes by learning a sequence of optimal RIS configurations. The loss function in \eqref{lossu} encourages the model to design a sequence of $T$ RIS configurations that minimize the localization error in the final stage. After training, the network generalizes across the deployment area and dynamically generates \acrshort{ris} configuration vectors based on the recurring received signals for any user and $S_{\text{RIS}}$, regardless of the location of the user within the coverage area. Notably, the loss function for the location of the user focuses solely on the estimation error in the final stage $T$. This is good since it gives the network the freedom to design the \acrshort{ris} configuration strategy at all \( T \) measurement stages. 

\section{Numerical Results}
\label{results}
\subsection{Simulation Setup}
The simulation parameters are summarized in Table~\ref{simulationparams}. We consider two scenarios in the experiments, \acrshort{siso} and \acrshort{mimo} to account for the impact of the antenna array at transceiver nodes in addition to \acrshort{ris} in \acrshort{rse}. As shown in Figure \ref{fig1}, the location of \acrshort{bs} is fixed and the distributed \acrshort{ris} is installed at known locations. The unknown user locations \( \mathbf{u} \) are uniformly distributed on the $x$-$y$ plane within the simulated enclosed area, arranged on a regular grid with $\lambda/2$ spacing and a known, fixed orientation. The pilot signal \( x^{(t)} \) is a known deterministic scalar fixed to unity. 
The $M$ \acrshort{so} trajectories are predefined a priori such that each \acrshort{so} starts at a known initial position and moves along a path with deterministic steps of maximum size $\lambda/4$, with random Gaussian offset in direction, $\Delta\theta \sim \mathcal{N}(0, \sigma_{\theta}^{2}), \quad \sigma_{\theta} = 0.052 \text{ radians},$ and position, $\Delta p \sim \mathcal{N}(0, \sigma_{p}^{2}), \quad \sigma_{p} = \lambda/20$, at each step. The positions of the central dipoles in each \acrshort{so} cluster are tracked in each simulation realization. The same base trajectories are used in training and testing, with independently sampled perturbations producing different realized paths.
The tunable dipole parameters for transceivers, environment (including \acrshort{so}), and \acrshort{ris} namely the resonance frequency (\( f_{\text{res}} \)), the charge term (\( \chi \)) and the absorptive damping term (\( \Gamma_L \)), are configured as in Table~\ref{simulationparams} \cite{9856592}.



\begin{table}[t]
    \caption{Numerical Results Parameters}
    \centering
    \begin{tabular}{|c|c|}
        \hline
        \textbf{Label} & \textbf{Dimension} \\ 
        \hline
        \multicolumn{2}{|c|}{\textbf{Simulation Parameters}}\\
        \hline
        $M$ & 4\\ 
        \hline
         $N_{\text{BS}}$& 
         \{1,4\}\\
        \hline
        $N_{\text{U}}$& 
         \{1,4\}\\
       \hline
        $N_{\text{RIS}}$& 
         {20,60,100}\\
        \hline   
        Transceivers dipole parameters & $f_{\text{res}} = 1$, $\chi = 0.5$, $\Gamma_L = 0$ \\
        \hline
        Transceivers dipole spacing &\(\lambda/2 \)\\
        \hline
        Environment and \acrshort{so} dipole parameters & $f_{\text{res}} = 10$, $\chi = 50$, $\Gamma_L = 10^4$ \\
        \hline
        Environment and \acrshort{so} dipole spacing &\(\lambda/4 \)\\
        \hline
        RIS dipole parameters & $f_{\text{res}} \in \{0.8, 1.2\}$, \\ & $\chi = 0.2$,$\Gamma_L = 0.03$\\
        \hline
        RIS dipole spacing &\(\lambda/2 \)\\
        \hline
        \multicolumn{2}{|c|}{\textbf{Parameters of the \acrshort{bilstm} Network}}\\
        \hline
        $B$ & [32,64,128]\\
        \hline
        $N_L$ & [2,4,6]\\ \hline
        $d_s$ & [25,35,45] \\ \hline
        $\eta$ & [0.001,0.01,0.005,0.05]\\
        \hline
         Maximum epochs & 200\\
        \hline
        $L_n$&4\\
        \hline
    \end{tabular}
    \label{simulationparams}
\end{table}
\subsection{Proposed Approach and Baseline Approaches}
The proposed \acrshort{bilstm} approach for adaptive RIS configuration design via sensing of \acrshort{rse} for user localization is implemented using the parameters described in Table~\ref{simulationparams}, which are optimized using \acrshort{bo}. 
For training, the network is trained on 1,806,000 samples over 200 epochs. The training data for \acrshort{bisen} comprises of channel response measurements at $S_{\text{RIS}}$ and for \acrshort{biaruln} channel response measurements at the user alongside the outputs of \acrshort{bisen}. We need channel responses to generate the received pilots. These measurements are generated by moving \acrshort{so} across arbitrary user locations in the simulation. The training labels for \acrshort{bisen} are of dimension $2M$ which includes the two-dimensional true \acrshort{so} locations, whereas the labels for \acrshort{biaruln} correspond to the two-dimensional true user locations. We partitioned the dataset into 70\% for training, 15\% for validation, and 15\% for testing.
Once the BiLSTM model is trained, it autonomously determines a sequence of \acrshort{ris} configurations utilizing recurring sequential scattering estimation and user received pilot measurements, enabling accurate localization of any user within the coverage area. We evaluated the localization accuracy of our proposed approach in comparison with the following baseline approaches. 
\begin{enumerate}
    \item \textit{Random RIS Configurations}: The sequence of RIS configurations at each time instant are randomly chosen and do not adapt dynamically based on sensing of the scattering in \acrshort{rse}. Randomness in configurations introduces diversity in the observed signals, which is leveraged during training. A \acrshort{dnn} with dimensions [200, 200, 200, 2] is used to map the magnitude and phase of the user response received over \( T \) time frames, \( [\textbf{y}^{(t)}]_{t=0}^{T-1} \) to predict the location of the user by minimizing the loss in \eqref{lossu}. 
    
    \item \textit{Codebook based Approach}: This employs a codebook-based \acrshort{ris} configuration design such that a BiLSTM model is trained to learn optimal \acrshort{ris} configurations corresponding to various \acrshort{so} states, as in \cite{10765779}. During inference, the BS estimates the current environment and selects the closest RIS configuration from the precomputed codebook to optimize the accuracy of the user localization quantized by the loss in \eqref{lossu}. 

    \item \textit{DNN based Scattering Estimation}: Here, we use \acrshort{biaruln} for user localization, but replace the sequential modeling of \acrshort{bisen}  with a \acrshort{dnn} with dimensions [200, 200, 200, $2M$] to estimate \acrshort{so} location from the sensed signals \( [\textbf{y}_s^{(t)}]_{t=0}^{T-1} \) at $S_{\text{RIS}}$.
    
    \item \textit{\acrshort{biaruln} only}: This model uses \acrshort{biaruln} for user localization and \acrshort{ris} configuration design, but does not take into account scattering in the environment, according to \cite{10373816, 10279094}.
\end{enumerate}
All baseline models have been evaluated for the MIMO setting, while the proposed approach is tested in both \acrshort{siso} and \acrshort{mimo} scenarios.

\subsection{Results}
\begin{figure}[t!]
    \centering
    \includegraphics[width=1\linewidth]{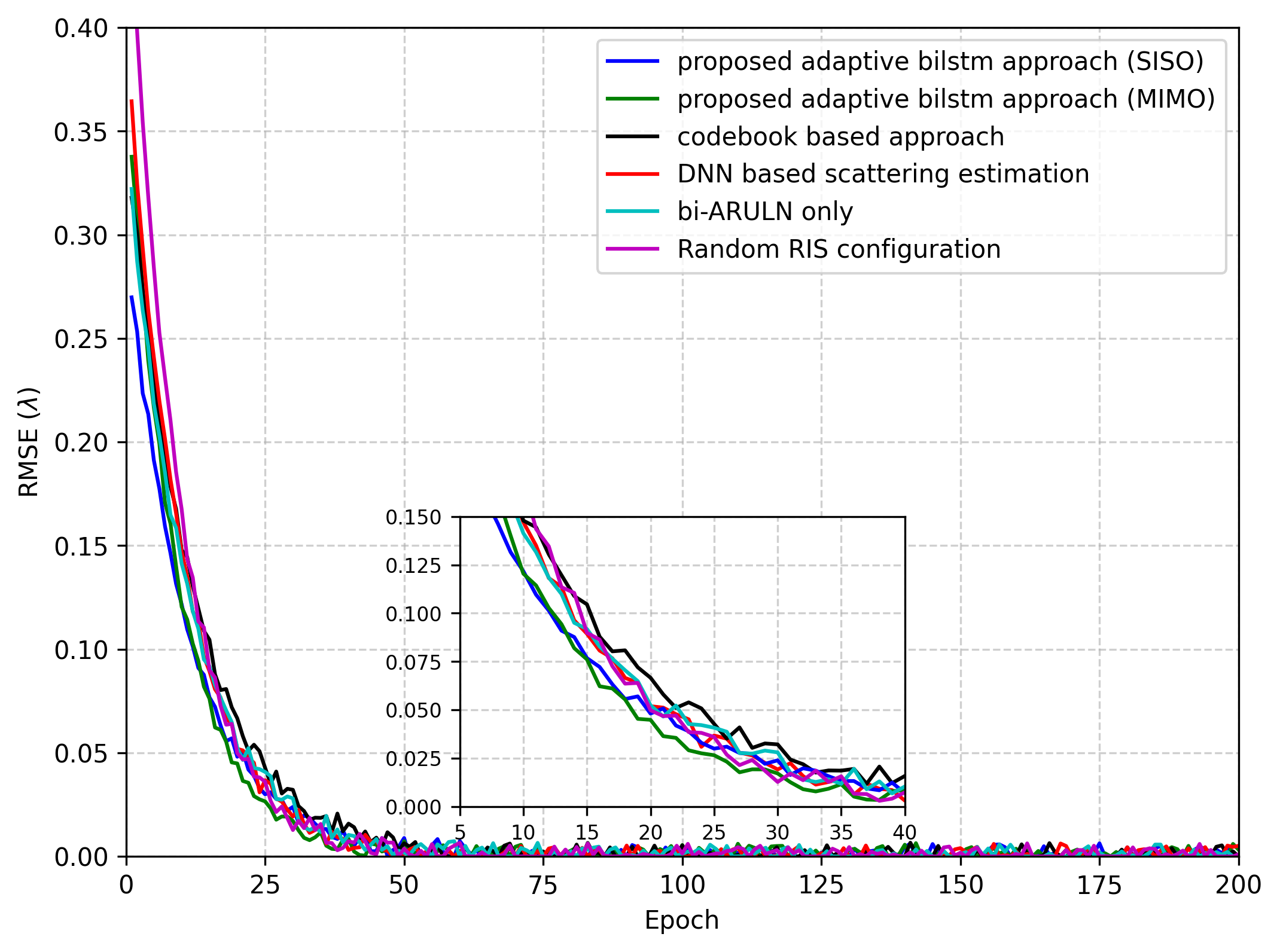}
    \caption{RMSE loss curves for proposed and baseline approaches for 200 evaluation epochs.}
    \label{fig4}
\end{figure}

The results in Figure~\ref{fig4} show the user localization \acrshort{rmse} loss curves for various approaches, demonstrating fast convergence across all methods. The loss function exhibits a rapid decline within the first 25 epochs, indicating efficient learning and adaptation of model parameters. The loss stabilizes near zero after epoch 50, indicating effective training with no signs of overfitting or divergence. Smooth convergence confirms that the models generalize well and that the training process was efficiently optimized. 

\begin{figure}[t!]
    \centering
    \includegraphics[width=1\linewidth]{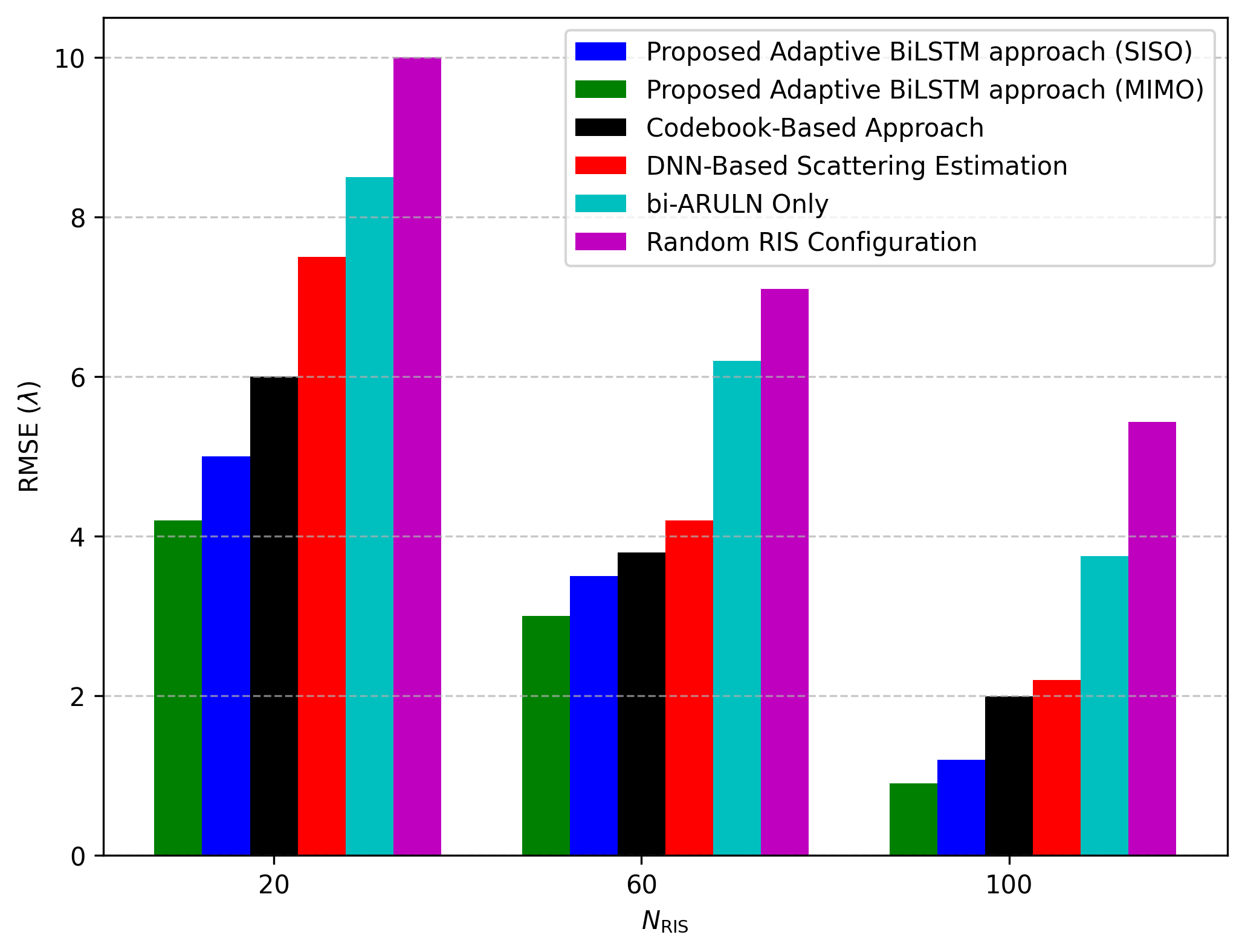}
    \caption{RMSE of user localization versus $N_{\text{RIS}}$, for $T$ = 21 and SNR = -20 dB.}
    \label{fig5}
\end{figure}

To evaluate the performance of the developed approach against the benchmarks, we first examine the effect of \acrshort{ris} size $N_{\text{RIS}}$ on localization accuracy, as shown in Figure~\ref{fig5}. The results show that increasing $N_{\text{RIS}}$, consistently reduces \acrshort{rmse} across all approaches, confirming the benefit of large \acrshort{ris} sizes on localization performance in dynamic \acrshort{rse}. This performance gain is expected, as a larger \acrshort{ris} provides more spatial information. The proposed approach achieves the lowest RMSE across all \acrshort{ris} sizes, indicating its effectiveness in adaptively optimizing RIS configurations for accurate localization compared to the baseline approaches. In particular, there is a 79\% reduction in \acrshort{rmse} as \( N_{\text{RIS}} \) increases from 20 to 100 for the \acrshort{mimo} network, and a 76\% reduction for the \acrshort{siso} network. The \acrshort{mimo} network consistently outperforms \acrshort{siso} due to its greater spatial diversity. All other methods show meaningful but relatively smaller drops, that is, below 70\%, in error. The proposed \acrshort{bilstm} approach remains scalable across various \acrshort{ris} sizes. Adapting the localization process for different \acrshort{ris} dimensions requires only a minor modification in the architecture of the neural network, specifically, by modifying the output size of the neural network in the \((L+1)\) layer following the \acrshort{bilstm} cells. The findings emphasize the importance of the adaptive \acrshort{ris} design in \acrshort{ris}-assisted and dynamic \acrshort{rse}. It is also observed that incorporating antenna array at the transceiver nodes further enhances the localization accuracy due to increased spatial diversity. 

\begin{figure}[t!]
    \centering
    \includegraphics[width=0.98\linewidth]{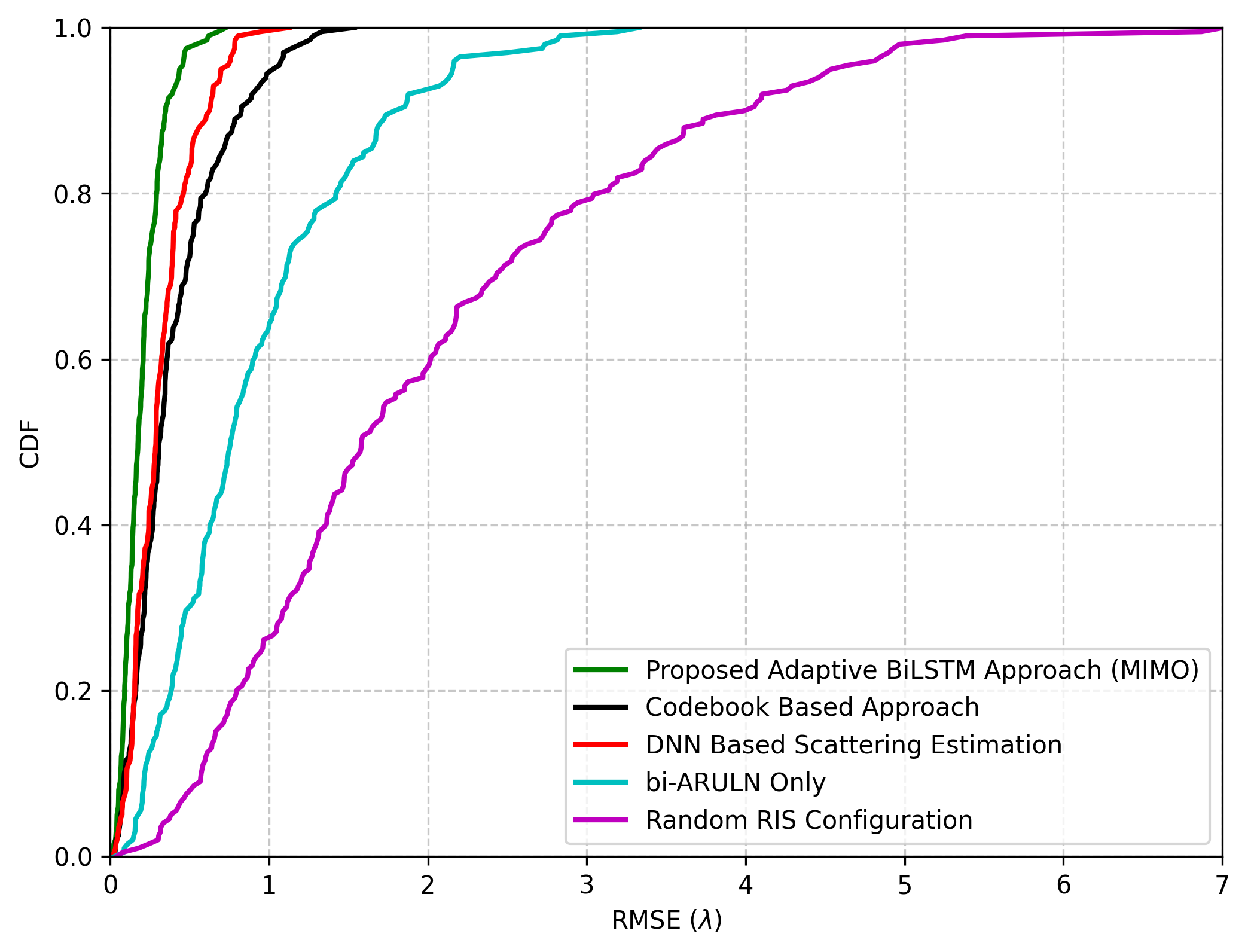}
    \caption{\acrshort{cdf} of proposed and baseline approaches versus RMSE of user localization for $N_{\text{RIS}} = 100$, $T = 21$ and SNR = 30 dB.}
    \label{fig6}
\end{figure}

Next, we examine \acrfull{cdf} of the user localization error from \eqref{lossu} for the proposed and benchmark approaches, as shown in Figure~\ref{fig6}. The proposed adaptive BiLSTM approach exhibits the most favorable distribution, with most localization errors concentrated at lower \acrshort{rmse} values, indicating high accuracy and reliability. The codebook-based approach and DNN-based scattering estimation follow closely, demonstrating relatively good performance but with slightly higher RMSE variability. In contrast, the bi-ARULN method and random \acrshort{ris} configuration show notably poorer performance. Their \acrshort{cdf} curves rising more gradually indicate higher and more variable errors. The random \acrshort{ris} configuration, in particular, exhibits the worst performance, with a large fraction of instances experiencing high RMSE values. These results confirm that adaptive, learning-based \acrshort{ris} configuration approaches offer substantial improvements in localization accuracy, while non-adaptive approaches or adaptive approaches not taking scattering into account suffer from greater uncertainty in dynamic \acrshort{rse}.
\begin{figure}[t!]
    \centering
    \includegraphics[width=0.98\linewidth]{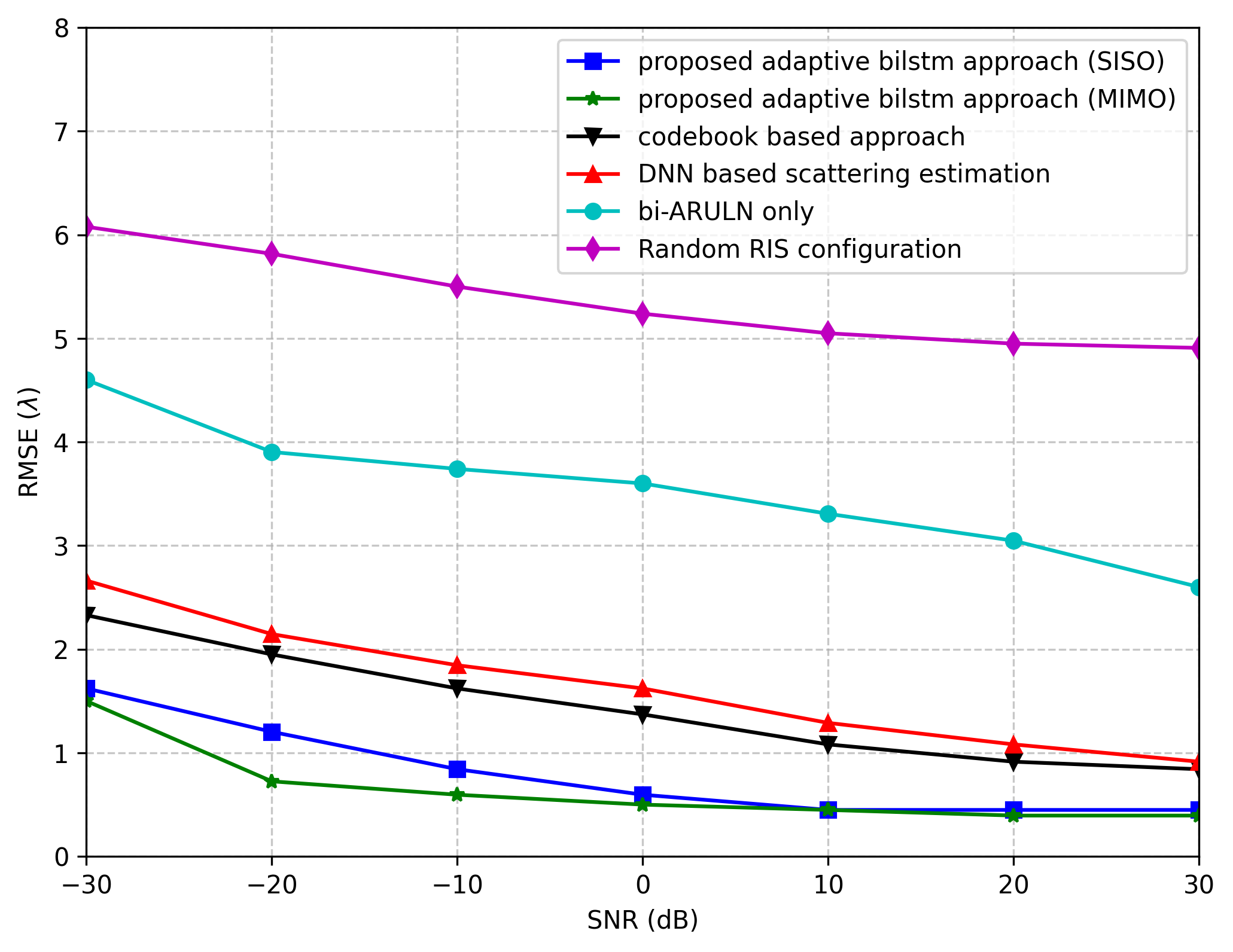}
    \caption{RMSE of user localization versus the SNR for $N_{\text{RIS}} = 100$ and \textit{T} = 21.}
    \label{fig7}
\end{figure}

We now examine the user localization performance against the \acrshort{snr}, i.e., $\text{SNR}_{\text{dB}} = 10 \log_{10} \left(P_{BS}\mathbb{E}\left[ ||\textbf{H}_{BS-U}||^2 \right]/ \sigma^2\right)$ in Figure~\ref{fig7}, which plots the  accuracy of the user localization versus \acrshort{snr}, where $P_{BS}$ is \acrshort{bs} transmit power. The results indicate that increasing \acrshort{snr} significantly improves the accuracy of the localization in all approaches, with varying degrees of effectiveness. The proposed approach consistently outperforms other benchmark approaches with nonadaptive designs or ones that do not take into account the scattering in \acrshort{rse}. This demonstrates its superior ability to refine \acrshort{ris} configurations using historical and current channel observations, ensuring its adaptive nature against other approaches. Both the \acrshort{siso} and \acrshort{mimo} networks in the proposed approach exhibit a consistent improvement with increasing \acrshort{snr}. MIMO generally outperforms \acrshort{siso} due to spatial diversity, particularly evident at low \acrshort{snr}s where robustness is critical. The narrow gap between \acrshort{siso} and \acrshort{mimo} performance at the lowest and highest \acrshort{snr} values likely reflects regimes where the estimator is dominated by noise at low SNR or by model and system limitations at high SNR. In both extremes, additional antennas provide little exploitable spatial information, and the benefits of MIMO become visible mainly in the mid-SNR range where spatial features can be effectively resolved.

The codebook-based approach, which utilizes pre-learned \acrshort{ris} configurations, shows moderate performance since it has limited adaptability to varying propagation environments, especially at lower \acrshort{snr}. 
The \acrshort{dnn} based scattering estimation approach performs worse than the codebook based and proposed approach,  but better than a method not taking scattering into account or with random \acrshort{ris} configuration. It shows a consistent drop with increasing \acrshort{snr}. This is because \acrshort{dnn}s can effectively model complex channel behaviors and compensate for non-ideal conditions in \acrshort{rse}. However, it does not fully capture sequential dependencies as efficiently as the \acrshort{bilstm}-based method.
The \acrshort{biaruln} only approach performs worse than deep learning-based approaches, mainly because it cannot effectively handle the scattering effects introduced by dynamic \acrshort{so} in \acrshort{rse}, even at higher \acrshort{snr}.

\begin{figure}[t!]
    \centering
    \includegraphics[width=1\linewidth]{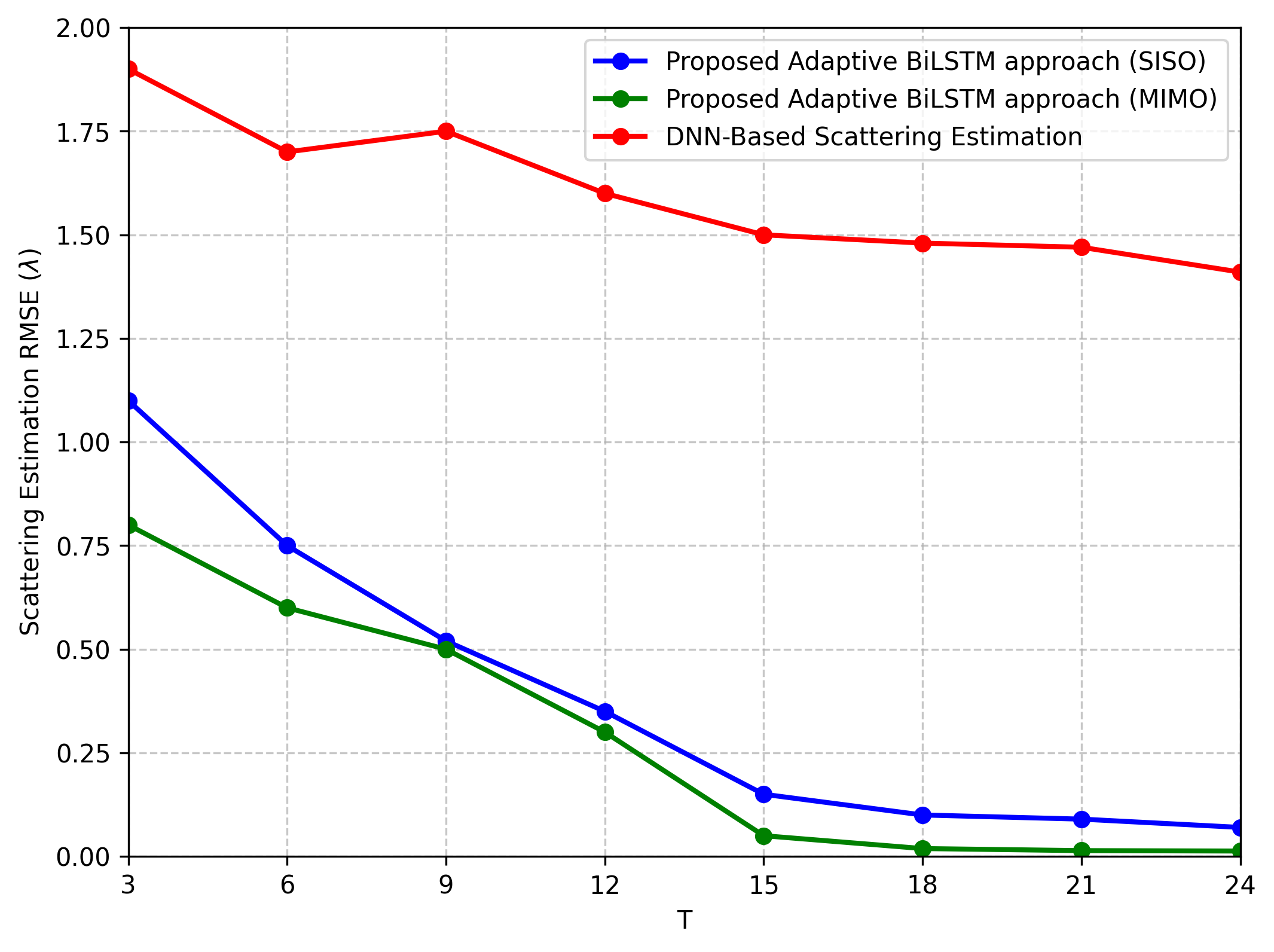}
    \caption{Scattering Estimation RMSE from \eqref{lossp} versus $T$ for \acrshort{snr} = 30dB.}
    \label{fig8}
\end{figure}

The random \acrshort{ris} configuration is not optimized for system requirements, leading to ineffective signal enhancement and poor localization accuracy such that its performance is largely insensitive to \acrshort{snr} changes. This underscores the necessity of accurate scattering estimation and well-designed adaptive \acrshort{ris} settings for improved localization. In general, the results highlight the superiority of adaptive learning-based approaches and the pivotal role of optimized \acrshort{ris} configurations in achieving high-precision user localization in \acrshort{ris}-assisted \acrshort{rse}.

In Figure~\ref{fig8} we study the scattering estimation error in \eqref{lossp}, accounting for cumulative RMSE in estimating $M$ \acrshort{so}, as a function of the number of time frames that is used for designing \acrshort{ris} configuration and estimating user location, highlighting the performance of different estimation techniques. The proposed \acrshort{bisen} approach, for both \acrshort{mimo} and \acrshort{siso} networks, outperforms the baseline by effectively leveraging sequential modeling for scattering estimation. As \( T \) increases, both the \acrshort{mimo} and \acrshort{siso} networks exhibit a significant reduction in RMSE, with MIMO achieving the lowest error due to enhanced spatial diversity and limited movement of \acrshort{so} along their trajectories, which allows the \acrshort{bisen} to leverage temporal correlations effectively. In contrast, the \acrshort{dnn}-based scattering estimation approach, shows consistently higher RMSE across all values of \( T \), as it lacks temporal modeling and thus fails to capture long-term scattering dependencies. Although its performance improves as \( T \) increases, it remains inferior to the \acrshort{bilstm}-based \acrshort{bisen} approach. In general, the performance gap between BiLSTM and DNN-based approaches highlights the importance of leveraging recurrent architectures for dynamic scattering estimation, reinforcing the benefit of adaptive, sequence-based learning in \acrshort{ris}-assisted localization in dynamic \acrshort{rse}.
\begin{figure}[t]
    \centering
    \includegraphics[width=0.98\linewidth]{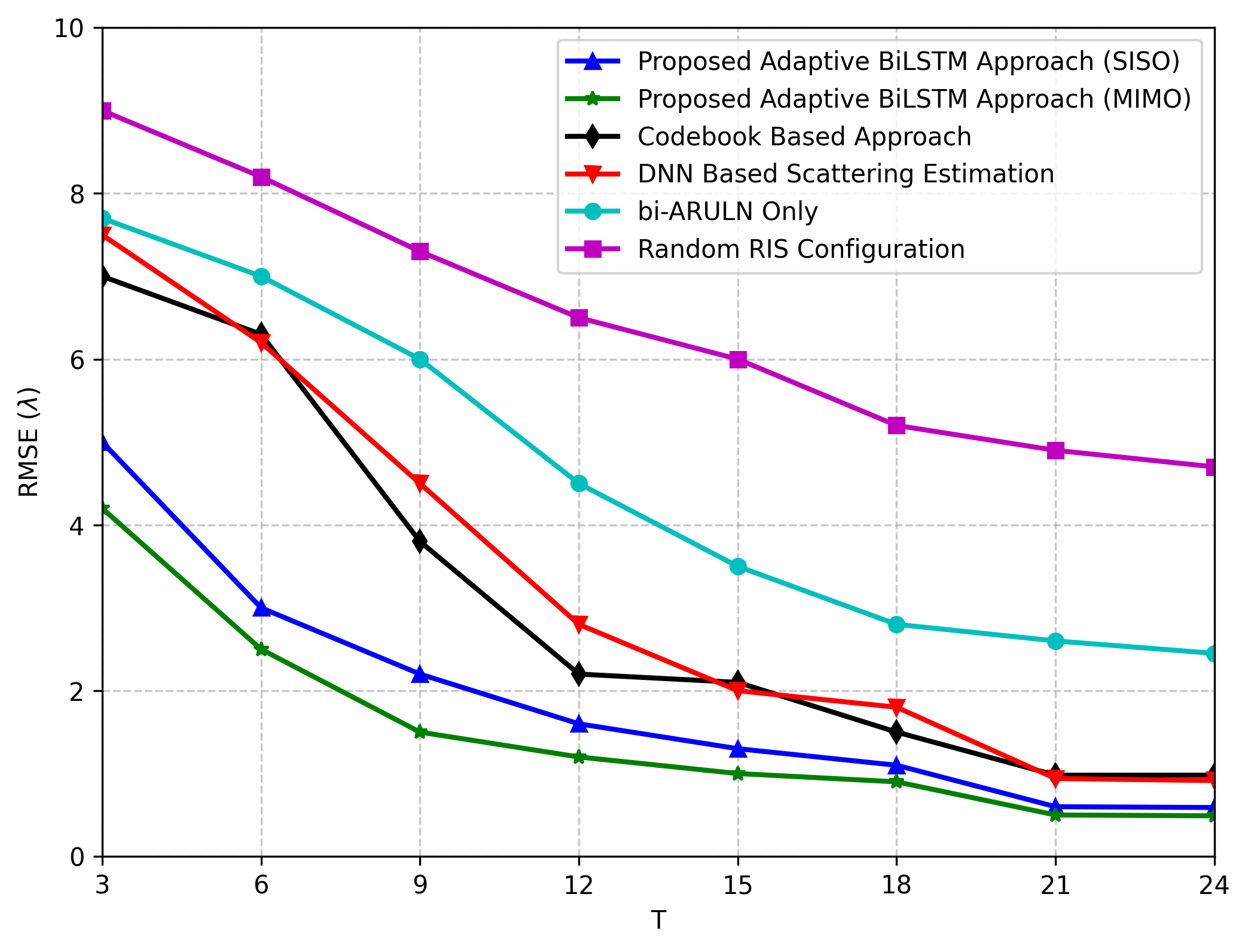}
    \caption{RMSE of user localization versus $T$ for $N_{\text{RIS}} = 100$ and SNR = 30 dB.}
    \label{fig9}
\end{figure}

Next, we examine the user localization error against varying time frames in Figure~\ref{fig9}. We observe that the devised approach performs consistently better than the benchmark approaches with the MIMO network giving slightly better performance than the \acrshort{siso} network. To demonstrate, observe that the devised approach with MIMO network at 6 time frames achieves a localization error of approximately 2.5$\lambda$ , which is comparable to or better than the error achieved by benchmark approaches at or after 12 time frames. This demonstrates a twofold reduction in the number of observations required to reach the same accuracy level. The performance gap is most pronounced at lower frame counts, highlighting the efficiency of the proposed method in rapidly extracting useful features from limited temporal information. In particular, the advantage of requiring fewer time frames translates into reduced localization latency. Although the gap narrows somewhat as the number of frames increases, the devised approach consistently maintains a margin over all benchmarks, indicating robustness across different operating conditions. The role of \acrshort{mimo} is also evident, as the richer spatial diversity in the \acrshort{mimo} network allows more reliable feature extraction, thus accelerating convergence to low-error regimes. This suggests that combining the proposed method with multi-antenna network configurations can further reduce latency and improve performance in practical deployments. From an application standpoint, these results imply that high localization accuracy can be achieved with fewer observations, which may lead to reduced latency, lower energy consumption, and diminished signaling overhead in comparison to the baselines. Such characteristics are particularly advantageous in power-limited or bandwidth-constrained scenarios, as well as in highly dynamic \acrshort{rse}s where long observation windows are impractical due to user or environmental mobility.   

\begin{figure}[t]
    \centering
    \includegraphics[width=0.98\linewidth]{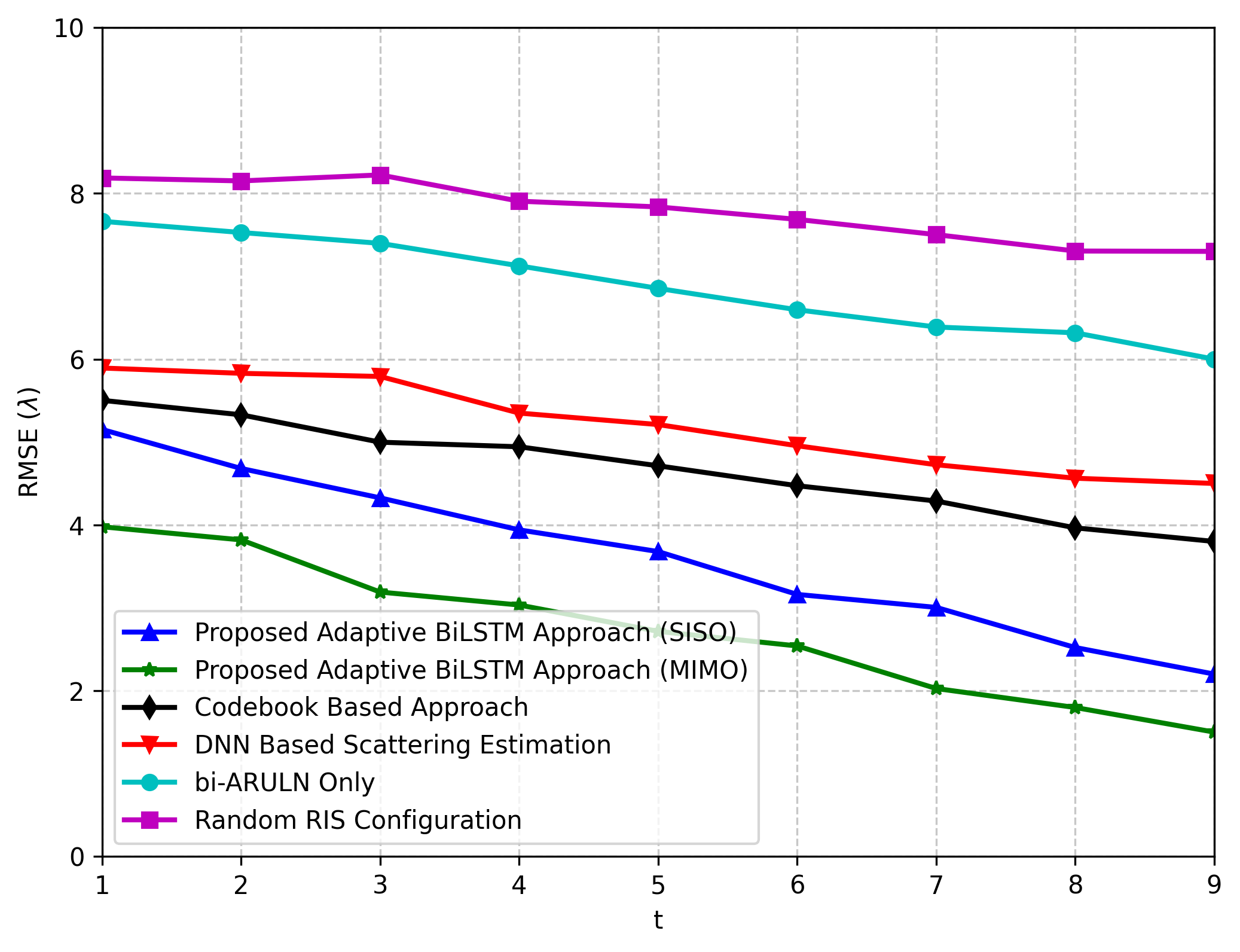}
    \caption{RMSE of user localization versus time frames $t$ up to the last time frame $T=9$ for $N_{\text{RIS}} = 100$ and SNR = 30 dB.}
    \label{fig10}
\end{figure}

We next plot the user localization \acrshort{rmse} as a function of the time stages $t \leq T$ in Figure~\ref{fig10}. 
To observe the \acrshort{rmse} at $t < T$, the localization process must be terminated before reaching the final stage $T$. In this case, the \acrshort{bilstm} network should still provide reliable position estimates at earlier stages of \acrshort{ris} configuration design and environment sensing. 
However, the loss function in \eqref{lossu} does not explicitly account for this requirement. To address this issue, we employ a weighted \acrshort{rmse} loss, defined as
$
\sqrt{\mathbb{E}[\sum_{t=1}^{T} \alpha_t \lVert \hat{\mathbf{u}}^{(t)} - \mathbf{u} \rVert^2]},
$
where $\hat{\mathbf{u}}^{(t)} = \ell_u([c_{u,f}^{(t)},c_{u,b}^{(t)}])$, and the weights satisfy $\sum_{t=1}^{T} \alpha_t = 1$. 
By assigning higher weights to earlier stages, the network is encouraged to reduce localization error even when early termination occurs.
\\
The results in Figure~\ref{fig10} illustrate that \acrshort{rmse} consistently decreases as $t$ increases, showing that each method progressively refines the \acrshort{ris} configuration with additional pilot observations. 
The proposed adaptive BiLSTM approach achieves the most significant improvements, consistent with the ability of recurrent models to capture and exploit temporal dependencies in the scattering environment.  
The use of multiple transceiver antennas provides richer information on the scattering geometry, thereby accelerating error reduction. In contrast, the DNN-based scattering estimator lacks explicit temporal memory and treats each frame independently, limiting its ability to exploit correlations over time. This explains its inferior performance relative to the BiLSTM-based methods, which effectively integrate information over multiple stages.
\\
The codebook-based approach performs moderately well, but is constrained by quantized \acrshort{ris} states. While it benefits from improved selections as the environment estimate stabilizes, its performance ceiling is ultimately dictated by the codebook resolution. Thus, although codebook-based methods are attractive for their simplicity, they cannot match the continuous adaptivity and robustness of recurrent approaches in \acrshort{rse}.
The adaptive \acrshort{biaruln}-only method relies solely on user-side signals without modeling scattering in the \acrshort{rse}, thus limiting its ability to design configurations suitable for environmental conditions. 
The random configuration strategy provides measurement diversity, but lacks directed adaptation, resulting in slow convergence and the highest error.
\\
A general trend across all methods is that performance gaps widen with $t$. The early stages capture only coarse information, whereas the later stages refine both the environment representation and the \acrshort{ris} design. This allows adaptive methods to accelerate improvements, while non-adaptive methods provide only limited gains. In the final stage, the proposed adaptive \acrshort{bilstm} with \acrshort{mimo} converges to the lowest error, demonstrating stable tracking of the \acrshort{so} and robust \acrshort{ris} configuration design in the considered \acrshort{rse}.
\\
From a practical standpoint, if the system is subject to stringent latency or pilot constraints, the proposed BiLSTM with \acrshort{mimo} should be prioritized where hardware resources allow, while the \acrshort{siso} variant offers an alternative with lower complexity. For applications requiring early termination, training with a weighted early-stage loss is particularly valuable, as it drives the network to place the \acrshort{ris} in informative states earlier, reducing RMSE even at small $t$.

\section{Conclusion}
\label{conclusion}
This paper presents a learning-based localization approach for \acrshort{ris}-assisted \acrshort{rse} to adaptively design the \acrshort{ris} configuration for accurate user localization by taking into account scattering in the environment. By leveraging \acrshort{ris}-assisted sensing and sequential measurement history, our proposed \acrshort{bilstm}-based model adaptively configures \acrshort{ris} elements to minimize user localization error. The integration of \acrshort{bisen} for scattering estimation and \acrshort{biaruln} for \acrshort{ris} configuration design allows the system to maintain context awareness and adaptively design \acrshort{ris} configuration for user localization. Simulation results validate that the proposed approach outperforms fixed-codebook, random, and adaptive designs that do not take scattering into account across varying \acrshort{ris} sizes and \acrshort{snr} conditions. The proposed approach also generalizes well to both \acrshort{siso} and \acrshort{mimo} networks. Given the reciprocity of the underlying channel model, the proposed approach can also be applied in an uplink configuration without modifying the core learning method. Future work may explore analytical benchmarking of the proposed solution, and quantizing the adaptive configuration space to enable deployment under practical bandwidth and storage constraints. Moreover, the extension of the proposed approach to the estimation of user orientation in addition to location and unstructured \acrshort{rse}s with unknown or varying numbers of \acrshort{so} and their mobility and different \acrshort{ris} deployment strategies may also be investigated.

\bibliographystyle{IEEEtran} 
\bibliography{references}

\vskip -2\baselineskip plus -1fil
\begin{IEEEbiography}[{\includegraphics[width=1in,height=1.25in,clip,keepaspectratio]{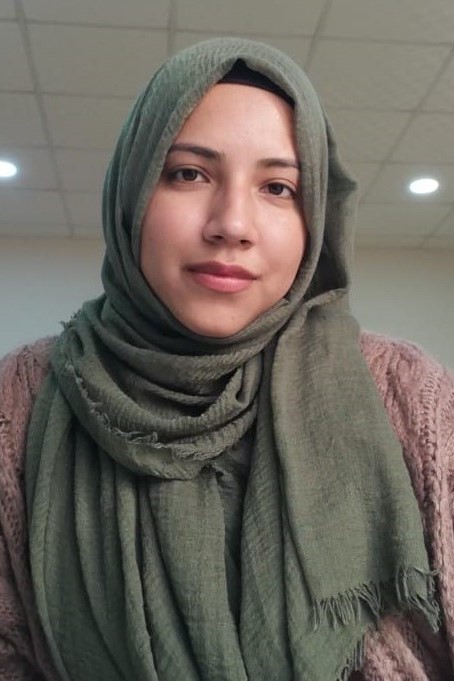}}]{Anum Umer} 
received the B.E. degree in electrical (telecommunication) engineering and the M.S. degree in electrical engineering from the National University of Science and Technology (NUST), Pakistan, in 2015 and 2017, respectively, where she was a Research Associate with the System Analysis and Verification Lab, School of Electrical Engineering and Computer Science in 2018.  During 2019-2022, she was a Research Engineer with the Research and Development Wing, NUST. She is currently pursuing her Ph.D. degree in Information and Communication Technology from the Thomas Johann Seebeck Department of Electronics, Tallinn University of Technology, Estonia. Her areas of research include wireless communication, localization, and sensing.
\end{IEEEbiography}
\vspace{-0.5cm}
\begin{IEEEbiography}[{\includegraphics[width=1in,height=1in,clip,keepaspectratio]{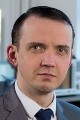}}]{Ivo M\"{u}\"{u}rsepp} 
was born in Tallinn, Estonia in 1980. He received bachelors and MSc degrees in telecommunications from Tallinn University of Technology (TUT), Estonia, in 2002 and 2004 respectively. Ph.D. degree in Telecommunications from Tallinn University of Technology was received in 2013. In 2002 he joined the Department of Radio and Communication Technology of TUT as a teaching assistant. Currently he works in the Thomas Johann Seebeck department of electronics as a senior lecturer. Has also been employed for 5 years by Cyber Command of Estonian Defense Forces. His research interest includes signal processing, radio frequency engineering, radio communication, and mobile positioning.
\end{IEEEbiography}
\begin{IEEEbiography}[{\includegraphics[width=1in,height=1.25in,clip,keepaspectratio]{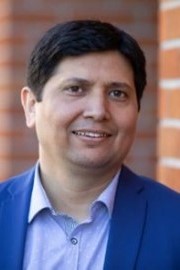}}]{Muhammad Mahtab Alam} 
(Senior Member, IEEE) received the M.Sc. degree in electrical engineering from Aalborg University, Denmark, in 2007, and the Ph.D. degree in signal processing and telecommunication from the INRIA Research Center, University of Rennes 1, France, in 2013. From 2014 to 2016, he was Post-Doctoral Research at the Qatar Mobility Innovation Center, Qatar. In 2016, he joined as the European Research Area Chair and as an Associate Professor with the Thomas Johann Seebeck Department of Electronics, Tallinn University of Technology, where he was elected as a Professor in 2018 and Tenured Full Professor in 2021. Since 2019, he has been the Communication Systems Research Group Leader. He has over 15 years of combined academic and industrial multinational experiences while working in Denmark, Belgium, France, Qatar, and Estonia. He has several leading roles as PI in multimillion Euros international projects funded by European Commission (Horizon Europe LATEST-5GS, 5G-TIMBER, H2020 5GROUTES, NATOSPS (G5482), Estonian Research Council (PRG424), Telia Industrial Grant etc. He is an author and co-author of more than 100 research publications. He is actively supervising a number of Ph.D. and Postdoc Researchers. He is also a contributor in two standardization bodies (ETSI SmartBAN, IEEE-GeeenICT-EECH), including ‘‘Rapporteur’’ of work item: DTR/ SmartBAN-0014. His research focuses on the fields of wireless communications–connectivity, mobile positioning, 5G/6G services and applications.
\end{IEEEbiography}

\end{document}